\documentclass[fleqn,usenatbib]{tjaa}

\usepackage[T1]{fontenc}
\usepackage{amsmath}		
\usepackage{amssymb}		
\usepackage[pdftex]{graphicx}	
\usepackage[utf8]{inputenc}
\usepackage[english]{babel}
\usepackage{newtxtext,newtxmath}
\usepackage{lipsum}
\usepackage{gensymb}

\title{Ç.Ü. Uzay Bilimleri ve Güneş Enerjisi Araştırma ve Uygulama Merkezi (UZAYMER): \\
I. Gözlem Koşulları ve Güncel Projeler}
\author[uzaymer]{
A. Solmaz,$^{1,2}$\thanks{E-mail: arif.solmaz@gmail.com}
N. Aksaker,$^{1,3}$
A. Akyüz,$^{1,4}$
Z. Kurt,$^{1,5}$
S. Allak,$^{1,6}$
Y. Aladağ,$^{1}$
\and
M. Karakılçık,$^{1,4}$
N. Emrahoğlu,$^{1,7}$
M. Emin Özel,$^{1}$
\\
$^1$Uzay Bilimleri ve Güneş Enerjisi Araştırma ve Uygulama Merkezi (UZAYMER), Çukurova Üniversitesi, 01330, Adana, Turkey\\
$^2$Uzay Gözlem, Uygulama ve Araştırma Merkezi, Çağ Üniversitesi, 33800, Mersin, Turkey\\
$^3$Adana Organize Sanayi Bölgesi Teknik Bilimler Meslek Yüksekokulu, Çukurova Üniversitesi, 01330, Adana, Turkey\\
$^4$Fizik Bölümü, Çukurova Üniversitesi, 01330, Adana, Turkey\\
$^5$Uzaktan Algılama ve Coğrafi Bilgi Sistemleri Anabilimdalı, Çukurova Üniversitesi, 01330, Adana, Turkey\\
$^6$Fizik Bölümü, Çanakkale Onsekiz Mart Üniversitesi, 17100, Çanakkale, Turkey\\
$^7$İlköğretim Fen Bilgisi Öğretmenliği, Çukurova Üniversitesi, 01330, Adana, Turkey\\
}
\date{Accepted XXX. Received YYY; in original form ZZZ}

\pubyear{2021}
\volume{2}
\issue{1}
\begin{document}
\label{firstpage}
\pagerange{\pageref{firstpage}--\pageref{lastpage}}
\maketitle

\begin{abstract}
Çukurova Üniversitesi Uzay Bilimleri ve Güneş Enerjisi Araştırma ve Uygulama Merkezi (Ç.Ü. - UZAYMER) 1991 yılında kurulmuştur. Bu çalışmada, UZAYMER'de sürdürülen astronomi, astrofizik ve güneş enerjisi alanlarındaki güncel araştırmalar ile merkezin sahip olduğu astronomik altyapı hakkında bilgi verilmektedir. 
Bununla birlikte, AstroGIS veri tabanından derlenen uzun dönem (20 yıllık) ve meteoroloji istasyonundan alınan orta dönem (7 yıllık) ölçümler ile UZAYMER'in astrometerolojik durumu değerlendirilmektedir. Gözlemevinde sürdürülen bilimsel gözlemlere ait astronomik sönümleme katsayıları, görüş koşulları ve ışık kirliliği ölçümleri sunulmaktadır. 50 cm ayna çapına sahip UZAYMER Teleskobu'nun (UT50) gözlem limitleri hesaplanarak gözlem performansı değerlendirilmiştir. İyileştirilen teknik altyapısı ve artan insan kaynağı ile UZAYMER hem ulusal hem de uluslararası bilimsel iş birliklerini sürdürebilecek bir seviyeye ulaşmıştır. Ayrıca UZAYMER, yıllık açık gece sayısı, uygun hava koşulları ve karanlık gökyüzü değerleri ile 2141 gözlemevi arasında Astronomik Gözlemevleri Uygunluk İndisi (SIAS) ölçeğinin A kategorisinde 972. sırada yer almaktadır.
\end{abstract}

\begin{abstract}
Çukurova University Space Sciences and Solar Energy Research and Application Center (UZAYMER) was founded in 1991. In this work, we present current research in the fields of astronomy, astrophysics and solar energy and the astronomical infrastructure of the center. In addition, the astrometrological status of UZAYMER was investigated using long-term (20 years) data from the AstroGIS database and moderate-term (7 years) measurements taken from the meteorological station. Astronomical extinction coefficients, seeing conditions and light pollution measurements of UZAYMER were also obtained. Furthermore, the observation limits were determined for the UZAYMER 50 cm telescope (UT50) and the observation efficiency was evaluated.
UZAYMER has reached a level that can support both national and international scientific collaborations with its improved technical infrastructure and increased human resources. 
With its annual number of clear nights, favorable weather conditions and dark sky values, UZAYMER ranks 972 in Group A of the Suitability Index for Astronomical Sites (SIAS) among the 2141 observatories.
\end{abstract}


\section{Giriş}

Çukurova Üniversitesi (Ç.Ü.) Uzay Bilimleri ve Güneş Enerjisi Araştırma ve Uygulama Merkezi (UZAYMER)'in kuruluşu, Fen Edebiyat Fakültesi (o zamanki adıyla, Temel Bilimler Fakültesi) kurucu dekanlık görevini yürüten Prof. Dr. Hakkı Ögelman'ın, "Adana koşullarında iklimlendirme işlevleri olan bir 'Güneş Evi' kurulması" projesi (1981) ile başlamıştır. Bu proje, yenilenebilir enerji kaynaklarını (Güneş enerjisi, biyogaz, vb.) kullanarak ısıtma, soğutma (iklimlendirme), sıcak su ve yakıt gibi girdilerin olabilirliğini denetlemek amacıyla oluşturulmuştur. Bu alanda elde edilen ilk başarılı sonuçlar yayınlanmıştır \citep{Altun1985}.

1990 yılında Ç.Ü. Fizik bölümüne katılan Prof. Dr. Mehmet Emin Özel bu yapı ile ilgilenerek Çukurova Üniversitesi, Bonn Max Planck Enstitüsü ve Humboldt Vakfı'nın desteklerini alan projesi ile Güneş Evi'nin Uzay Bilimleri ve Güneş Enerjisi Araştırma ve Uygulama Merkezi'ne (UZAYMER-1991) dönüştürülmesini sağlamıştır. Başlangıçta, yüksek enerji astrofiziği, radyo astronomi, güneş enerjisi, biyogaz ve uydularla uzaktan algılama konularında yürütülen çalışmalar, daha sonra X-ışınları ve optik astronominin güncel konuları ve güneş enerjisine odaklanmıştır. Yürütülen ulusal ve uluslararası ortak çalışmalar ve projelerle birlikte, üretilen yayınlar ile yüksek lisans ve doktora tezlerinin gerçekleştirilmesi sağlanmıştır.

2010 yılında UZAYMER'e yeni yapılan bir gözlemevi binası ve kubbe ile 30-cm ayna çaplı teleskop kurulmuştur. Bu teleskopla her seviyede öğrenci için eğitim amaçlı gözlemlere başlanmıştır. 2017 yılında, kayar çatılı prefabrik bir gözlemevine yerleştirilen 50-cm ayna çaplı UZAYMER'in en büyük teleskobu \textbf{UT50} ile araştırma odaklı gözlemler başlatılmıştır. Uluslararası bir gözlem projesi kapsamında seçilen çift yıldız sistemlerinin fiziksel özellikleri UT50 gözlemleri ile belirlenmiştir \citep{2020arXiv200600528P, 2021NewA...8601571P}.

\section{Gözlem Araçları ve UT50 Gözlemevi Koşulları}
\begin{figure}
\begin{center}
\includegraphics[width=\columnwidth]{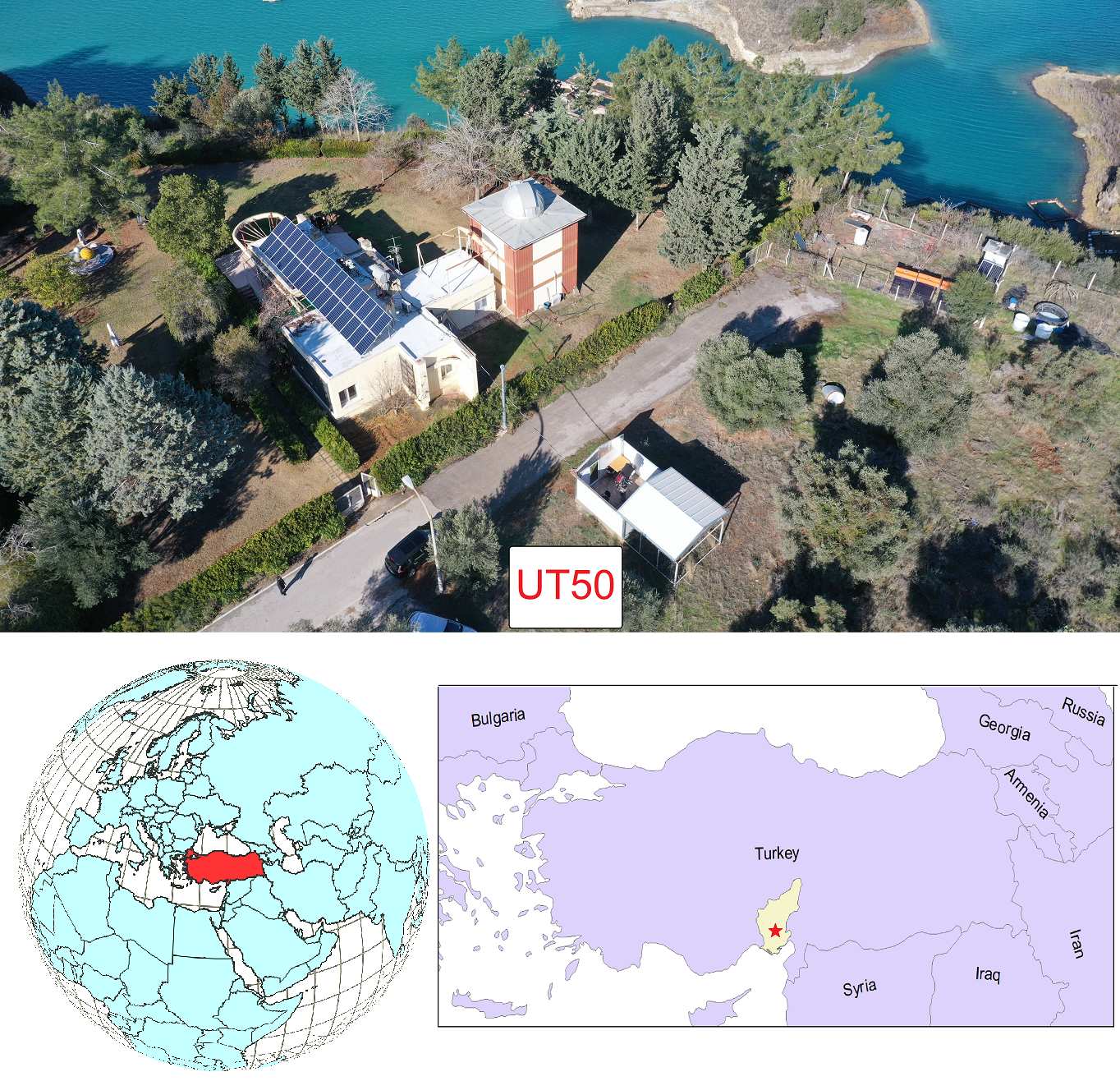}
\caption{%
UZAYMER'in Ç.Ü. yerleşkesi içindeki konumu. Güneş panellerinin bulunduğu yapı (paneller sonradan eklenmiştir) ve kubbeli binaya uzanan kısım, ilk Güneş Evi'nin bulunduğu mekandır. 2010'da yapılan Ç.Ü. Gözlemevi binası, Güneş Evi uzanımına bitişik kubbeli, 3 katlı yapıdır. Kayar çatılı, fotoğrafta üstü açık (çatısı kaydırılmış) prefabrik yapı, UT50'nin bulunduğu yapıdır. Üstte, Üniversite Yerleşkesini büyük ölçüde çevreleyen Seyhan Baraj Gölü görülmektedir. Altta, Türkiye'nin ve Adana'nın konumları daha geniş perspektiflerden verilmektedir. UZAYMER
$37\degree 03'20.36''K$ enleminde $35\degree 20'52.19''D$ boylamında ve deniz seviyesinden 130 m yükseklikte yer almaktadır.
}
\label{F:map}
\end{center}
\end{figure}

UZAYMER'de 1 adet 50 cm (UT50), 1 adet 30 cm ve 4 adet 20 cm çaplarında toplam 6 teleskop bulunmaktadır. Ancak bilimsel araştırmalar için en büyük ayna çapına sahip teleskop (UT50), amatör ve eğitim amaçlı faaliyetler için diğer teleskoplar kullanılmaktadır. UZAYMER'in konumu Şekil \ref{F:map}'de verilmektedir

\subsection{Gözlem Araçları}
UT50 teleskobu, Ritchey–Chrétien (RC) optiğine sahip ve GM4000 QCI Alman ekvatoryal kundaktan oluşmaktadır. 12 yuvalı FLI model filtre tekeri üzerinde standart UBVRI filtreleri ile gözlemler yapılmaktadır. QCI model odaklayıcı, filtre tekeri ile CCD arasında sıcaklık durumuna göre odak ayarı yapabilen bir kurguda bulunmaktadır. UT50'de geniş görüş alanına (30.4 açı dakikası) sahip 4096x4096 piksel (piksel ölçeği=0.891 açı saniye/piksel) FLI Proline PL16803 Monochrome model bir CCD kullanılmaktadır.

UZAYMER profesyonel bir gözlemevi altyapısına sahiptir. İnternet altyapısı elektrik kesintisinden bağımsız olup bu kesintiler gözlemleri etkilememektedir. Bunu sağlamak için UT50 teleskop binası 2 kVA'lık bir güç birimi (UPS) ile desteklenmektedir. Anlık veri indirme hızı $\sim$100 Mbps'dir. Teleskoba bağlı bilgisayarda (Win8) dış kullanım için güvenlik duvarı düzenlemeleri yapılmıştır. Uzaktan denetim masaüstü programları (Temaviewer, Anydesk vb) ile UT50 uzaktan kullanıma uygun hale getirilmiştir. 

\subsection{Gözlemevi Koşulları}
\subsubsection{Astronomik Görüş}

\begin{figure}
\begin{center}
\includegraphics[width=\columnwidth]{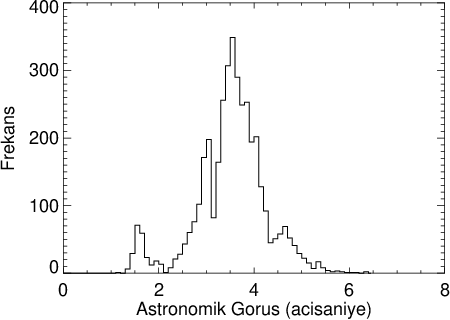}
\caption{%
UT50 verileri kullanılarak oluşturulan astronomik görüş değerlerinin histogramı. Astronomik görüş 1.5$''$, 3.5$''$ ve 4.5$''$ olmak üzere 3 farklı bölgede toplanmıştır. Histogram genel olarak ortalama astronomik görüş seviyesinde kümelenmiştir. 1.5$''$ civarında astronomik görüş kümelenmesi az olmakla birlikte UZAYMER'in zaman zaman iyi kalitede astronomik görüşe sahip olabildiği görülmektedir. Buradaki dağılım belli bir zaman aralığında kümelenmemiştir.
}
\label{F:seeing}
\end{center}
\end{figure}

UT50 sisteminde kurulu FLI CCD ile alınan, toplam 84 gecelik (2020) gözlem verisi kullanılarak astronomik görüş çalışması yapılmıştır. Bir IDL programı ile görüntülerde ortalama Yarı Yükseklikteki Tam Genişlik (FWHM) değerleri hesaplanmış ve CCD'nin piksel ölçeği ile çarpılarak astronomik görüş değerleri hesaplanmıştır. Filtre kaynaklı odak değişimlerinden etkilenmemek ve en fazla gözlem verisinden yararlanmak amacıyla R filtresinden alınan veriler kullanılmıştır. Toplamda astronomik görüş hesaplamaları için $\sim$ 4000 görüntü kullanılmıştır. Hesaplanan değerlerle oluşturulan astronomik görüş histogramı Şekil \ref{F:seeing}'de verilmiştir. Buna göre UZAYMER için medyan astronomik görüş 3.5 açı saniyesi olarak belirlenmiştir. Astronomik görüş en düşük 1.2 açı saniyesi ve en yüksek 6.4 açı saniyesi değerindedir. Astronomik görüş ölçümlerinden elde edilen bu değerler CCD görüntülerinin odak değişiminden etkilenmektedir. Ancak, astronomik görüş için kullanılan Diferansiyel Görüntü Takip Monitörü (Differential Image Motion Monitor - DIMM) ölçü aleti kullanılarak elde edilecek değerler bizim ölçümlerimizden daha düşük olacaktır.

\subsubsection{Gözlem Limitleri}

\begin{figure}
\begin{center}
\includegraphics[width=\columnwidth]{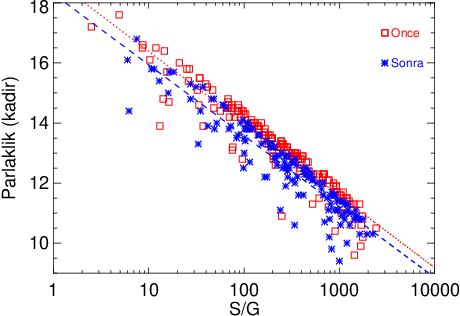}
\caption{%
UT50'nin birincil ayna yıkanmasından önce (kırmızı kareler) ve sonra (mavi yıldız) filtresiz gözlemlerinin sinyal gürültü oranına göre değişimi. Ölçüm değerlerine en iyi uyumu veren fitlerin denklemleri kırmızı nokta ve mavi kesikli çizgi için sırasıyla y=-2.4x+18.79 ve y=-2.4x+18.30 olarak belirlenmiştir. Çizgilerin seviyeleri arasında $\sim$0.5 kadir fark görülmektedir.
}
\label{F:mirror}
\end{center}
\end{figure}

UT50 aynalarının zaman içerisinde tozlandığı belirlenmiştir. Bu nedenle birincil aynanın yıkanmasına karar verilmiştir. 28 Aralık 2020 tarihinde saf su kullanılarak birincil ayna temizlenmiştir. Bu işlemden bir gece önce ve bir gece sonra 120 s poz süresiyle filtresiz NGC 103 açık küme gözlemi yapılarak kümedeki yıldızların parlaklık değişimleri incelenmiştir. Veri indirgemeleri sonrasında elde edilen sinyal gürültü oranına karşılık parlaklık grafiği Şekil \ref{F:mirror}'de verilmiştir. Bu grafiğe göre verilerde $\sim$0.5 kadir iyileşme gözlenmiştir.

\begin{figure*}
\begin{center}
    \begin{tabular}{@{}c@{}c@{}}
    \includegraphics[height=6.25cm]{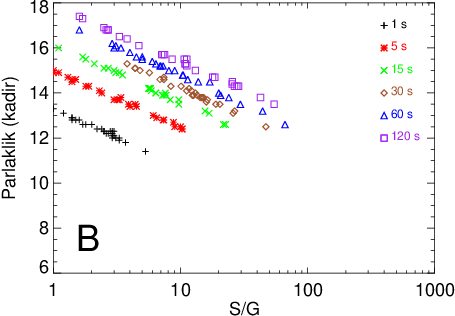} &
    \includegraphics[height=6.25cm]{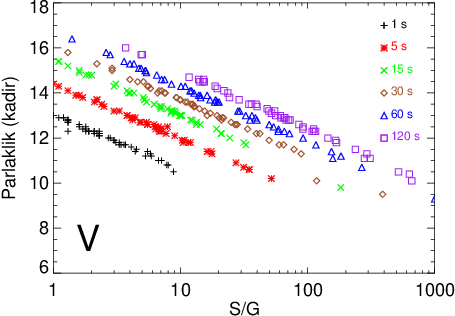} \\
    \includegraphics[height=6.25cm]{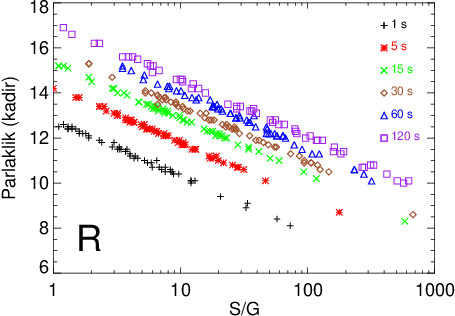} &
    \includegraphics[height=6.25cm]{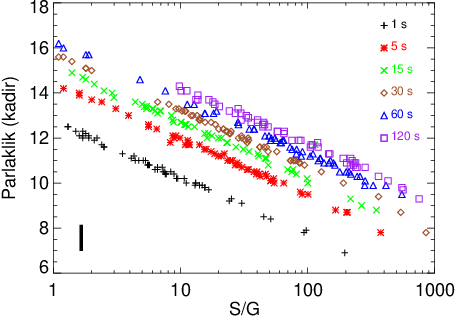} \\
    \end{tabular}
\caption{%
UT50'nin sırasıyla B, V, R ve I filtrelerindeki parlaklık değerlerine karşılık sinyal gürültü oranı değişimi. Grafiklerin sağ kısmında poz süreleri farklı semboller ile gösterilmiştir.
}
\label{F:limit}
\end{center}
\end{figure*}

UT50'nin gözlem limitlerinin belirlenmesi amacıyla NGC 103 açık kümesinin B, V, R ve I filtrelerinde gözlemleri yapılmıştır. Verilerin ön indirgemeleri AIJ (AstroImageJ) V3.2.0 ile fotometri işlemleri MaximDL V6.0 ile yapılmıştır. Şekil \ref{F:limit}'te B, V, R ve I filtrelerinde 1, 5, 15, 30, 60 ve 120 s poz süreleri için gözlem limitleri verilmiştir. Fotometrik sonuçlara göre 120 s poz süresinde 100 sinyal gürültü oranında (S/G) limit parlaklıklar B, V, R ve I filtrelerinde sırasıyla 12.9, 12.4, 12.0 ve 11.7 kadir olarak belirlenmiştir. Benzer şekilde 120 s poz süresinde 1 S/G için limit parlaklıklar B, V, R ve I filtrelerinde sırasıyla 17.9, 17.5, 17.2 ve 16.8 kadir olarak belirlenmiştir. Asteroit örtülme olayları ile Neptün-ötesi nesneler için 1 s poz süresinde beklenen $\sim$4-5 S/G oranında parlaklıklar sırasıyla 13.3, 13.0, 12.8 ve 12.7 kadir olarak hesaplanmıştır.

\subsubsection{Dönüşüm Katsayıları}

\begin{figure}
\begin{center}
\includegraphics[width=\columnwidth]{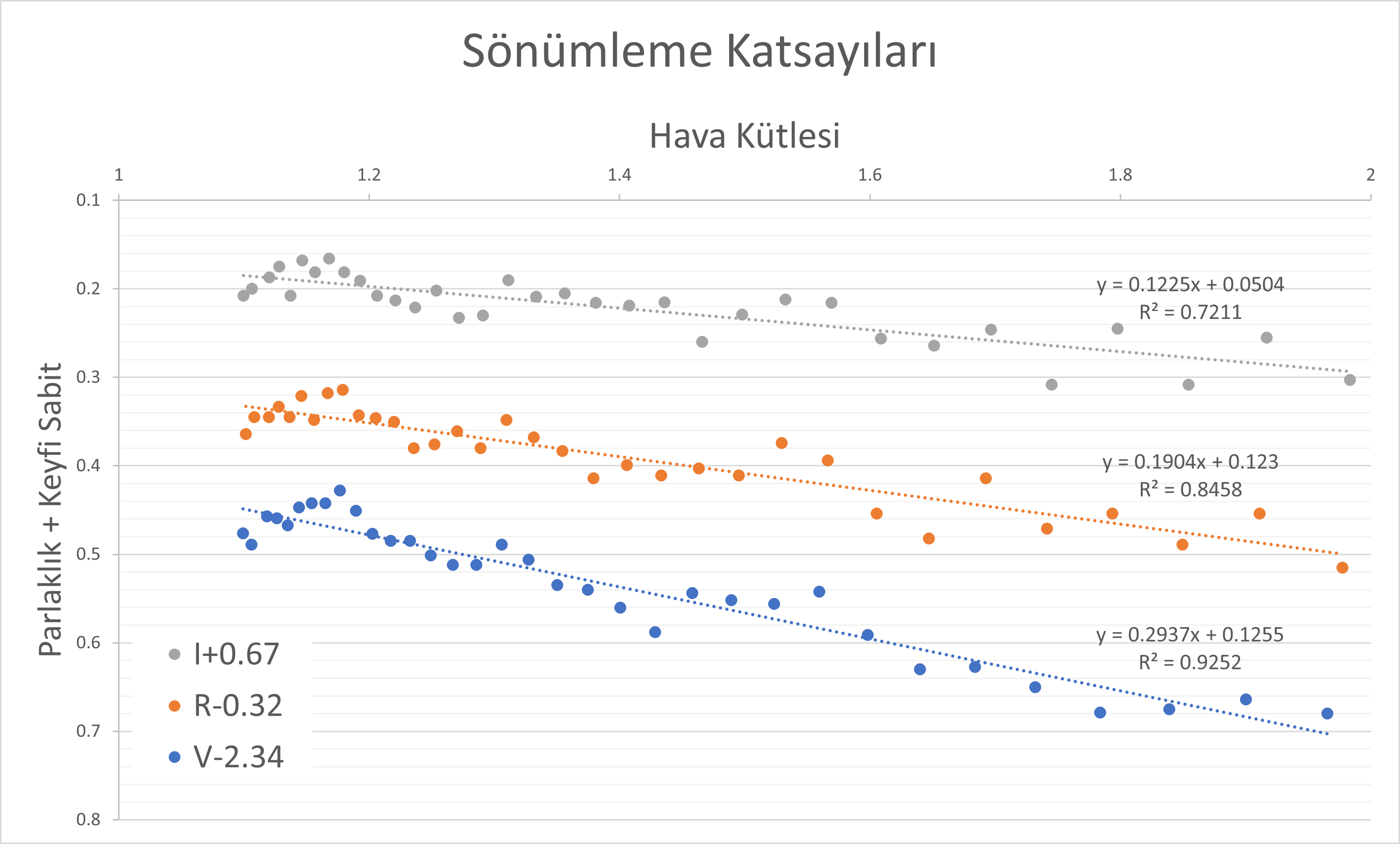}
\caption{%
UT50 ile yapılan standart yıldız gözlemlerinden elde edilen V, R ve I filtrelerindeki gecelik sönümleme katsayıları. Tüm gözlemlerde Landolt kataloğunda listelenen standart yıldızlar kullanılmış olup gecelik gözlemlerde ufuk yükseklikleri 30 ila 60 dereceye kadar nesnelerin hava kütlesi bilgileri kullanılmıştır. 
}
\label{F:sonumleme-katsayilari}
\end{center}
\end{figure}

UZAYMER'in atmosferik koşullarında alınan verilerin standart fotometrik sisteme dönüşüm katsayılarının belirlenmesi amacıyla V, R ve I bantlarında standart yıldızlar kullanılarak gözlemler yapılmıştır. Tüm gözlemlerde Landolt’un \citep{2009AJ....137.4186L} kataloğunda listelenen standart yıldızlar kullanılmıştır. Elde edilen sonuçlar Tablo \ref{T:tab_katsayılar}'de ve Şekil \ref{F:sonumleme-katsayilari}'te verilmiştir.

\subsection{UT50 Gözlemleri}

\begin{figure}
\begin{center}
\includegraphics[width=\columnwidth]{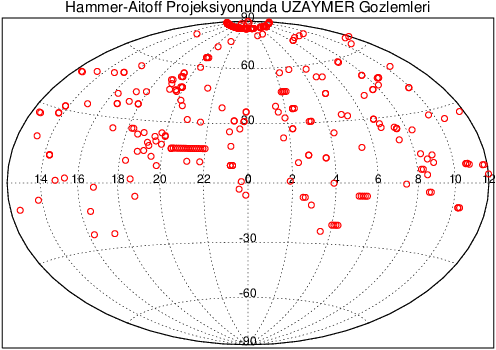}
\caption{%
UT50'nin gözlemlerinin (2299 adet) Hammer-Aitoff projeksiyonunda ekvatoryal koordinatlarda gösterimi. Gözlem koordinatları kırmızı halkalar şeklinde işaretlenmiştir. Kutup yıldızına yakın (grafikte üstteki yarım daire) ve 20-22 derece sağ açıklık - R.A. (dik açıklığı $\sim$20 derece) arasındaki sürekli çizgiler teleskobun park pozisyonunu göstermektedir.
}
\label{F:ut50_obs}
\end{center}
\end{figure}

\begin{figure}
\begin{center}
\includegraphics[width=\columnwidth]{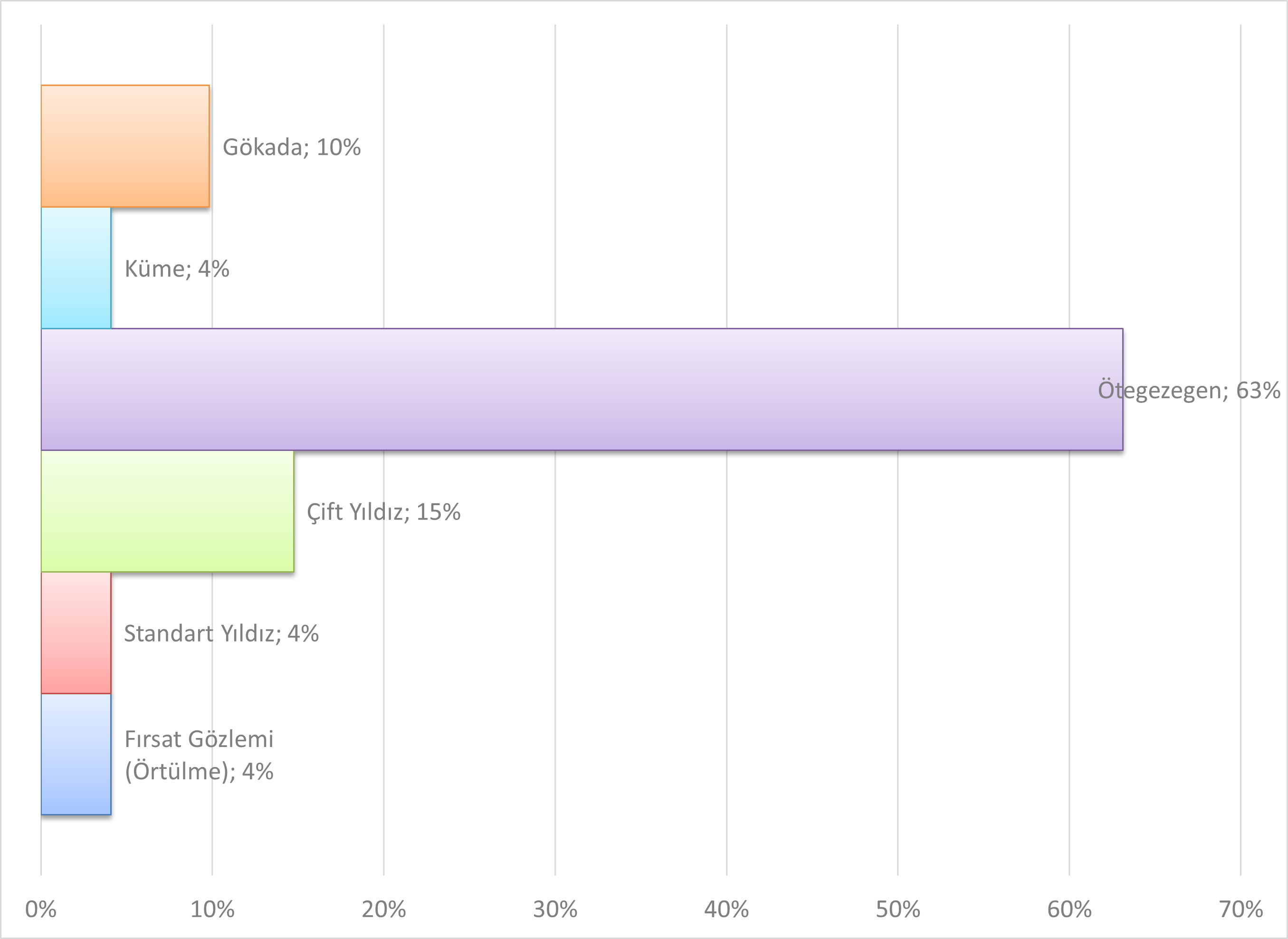}
\caption{%
UT50 ile gözlenen kaynak türlerinin dağılımı. Tür olarak en çok ötegezegen gözlemleri yapıldığından teleskop zamanı büyük oranda bu tür kaynakların geçiş gözlemlerine ayrılmaktadır.
}
\label{F:kaynak-turleri-yuzdesi-t50}
\end{center}
\end{figure}

UT50 ile ilk ışık 2013 yılında alınmıştır \citep{Nuri2013}. 2018 yılında Astronomi ve Astrofizik Anabili Dalı kurulana kadar teleskoplar bilimsel araştırmalarda etkin olarak kullanılamamıştır. 3 yıllık UT50 veri arşivi (65848 FITS dosyası) içerisindeki ötegezegen, standart yıldız, açık küme ve diğer gözlemlerden oluşan toplam 122 farklı kaynağın konumları Şekil \ref{F:ut50_obs}'da gösterilmiştir. Bu kaynakların nesne türlerine göre dağılımı Şekil \ref{F:kaynak-turleri-yuzdesi-t50}'de verilmektedir. UT50 teleskobu ile çoğunlukla lisansüstü tez çalışmalarında incelenen ötegezegenler gözlendiğinden ötegezegenler çoğunluktadır.

\section{Araştırma Konuları}
UZAYMER'de ulusal ve uluslararası iş birliklerine dayalı bilimsel projeler kapsamında X-ışın çiftleri, uzaktan algılama, güneş enerjisi, ötegezegenler, gezegenimsi bulutsular ve astronomik yer seçimi konularında araştırmalar yapılmaktadır.

\subsection{X-ışın Çiftleri}

Bir normal yıldız (kütle aktaran - Donör) ve bir sıkı cismin (beyaz cüce, nötron yıldız veya kara delik) oluşturduğu çift sistemlerde donör yıldızdan sıkı cisim üzerine madde yığılmasıyla X-ışın yayınımı oluşmaktadır. Bu sistemler, X-ışın çiftleri olarak tanımlanmaktadır. Galaktik ya da galaksi ötesi X-ışın çiftleri (XRB), X-ışın astronomisinin temel araştırma konuları arasındadır.
X-ışın çiftlerinin bir alt sınıfı olan Aşırı Parlak X-ışın kaynakları (APX) gökadaların merkez bölgesinde bulunmayan, ayrık kaynaklar olup X-ışını ışıtması L$_{X}$ > 10$^{39}$ erg s$^{-1}$ ile Galaktik XRB'ler ve Aktif Galaktik Çekirdekler arasında bir sınıf olarak tanımlanmaktadırlar. Bazı modeller, APX'lerin içerdiği yıldız kütleli (< 100 M$\odot$) kara delikler veya nötron yıldızları üzerine Eddington limitini aşan miktarda (süper-Eddington) yığılmayla gözlenen yüksek ışıtmaya ulaşabileceğini önermektedir \citep{2013MNRAS.432..506P,2013MNRAS.435.1758S,2013ApJ...778..163B,2017MNRAS.466L..48I,2017ARA&A..55..303K}. Bazı modellerde ise, APX'lerde bulunabilecek orta-kütleli  (100-10$^{4}$ M$\odot$) kara delikler üzerine Eddington limitinin altında (sub-Eddington) yığılma oranıyla bu ışımanın oluşabileceği belirtilmektedir \citep{1999ApJ...519...89C,2001ApJ...551L..27M,2015MNRAS.448.1893M}. M82 gökadasında bir APX X-2'den algılanan düzenli atımların (puls) keşfi ile bu çift sistemlerde süper-Eddington yığılma oranı ile yüksek ışıtmanın oluşma olasılığı önem kazanmıştır \citep{2014Natur.514..202B}. Şimdiye kadar, keşfedilen ve buna benzer atım kaydedilen APX sayısı altıdır \citep{2017Sci...355..817I,2016ApJ...831L..14F,2018MNRAS.476L..45C,2019MNRAS.488L..35S,2020ApJ...895...60R}.

Yeni nesil X-ışın uydularının (Chandra, XMM-Newton ve NuSTAR) gözlemleriyle bu kaynakların doğası hakkında önemli ipuçları elde edilmektedir. Chandra ve XMM-Newton gözlemevleri $\sim$0.2-10 keV ve NuSTAR ise 3-78 keV enerji aralığında fotonları algılamaktadır.
APX'lerin doğasını incelemek amacıyla X-ışın gözlemlerinin yanı sıra çoklu dalga boyu gözlemleri (optik, kırmızı ötesi, radyo) önemli bilgiler sunmaktadır. Bu kaynakların optik karşılıklarının belirlenmesi için Hubble Uzay Teleskobu (HST) ve 4m+ sınıfı teleskop verilerinden yararlanılmaktadır. 

\begin{figure}
\begin{center}
\includegraphics[width=\columnwidth]{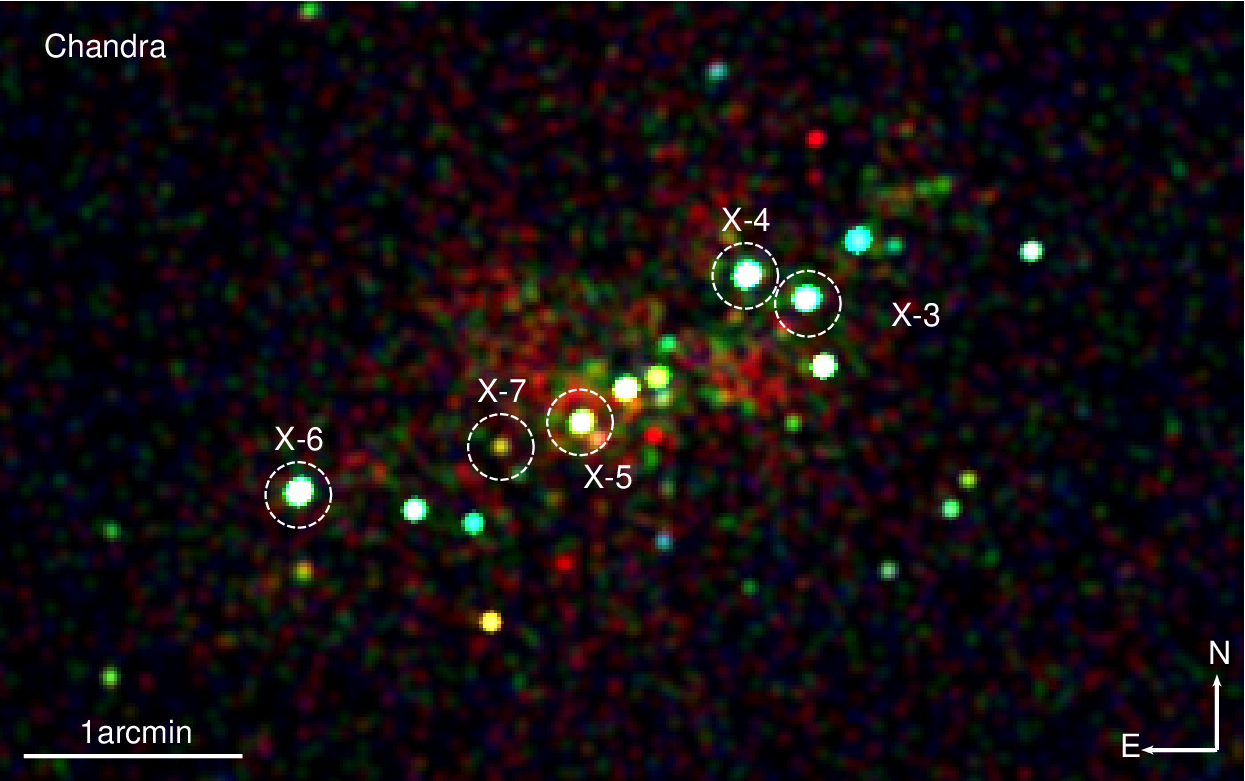}
\includegraphics[width=\columnwidth]{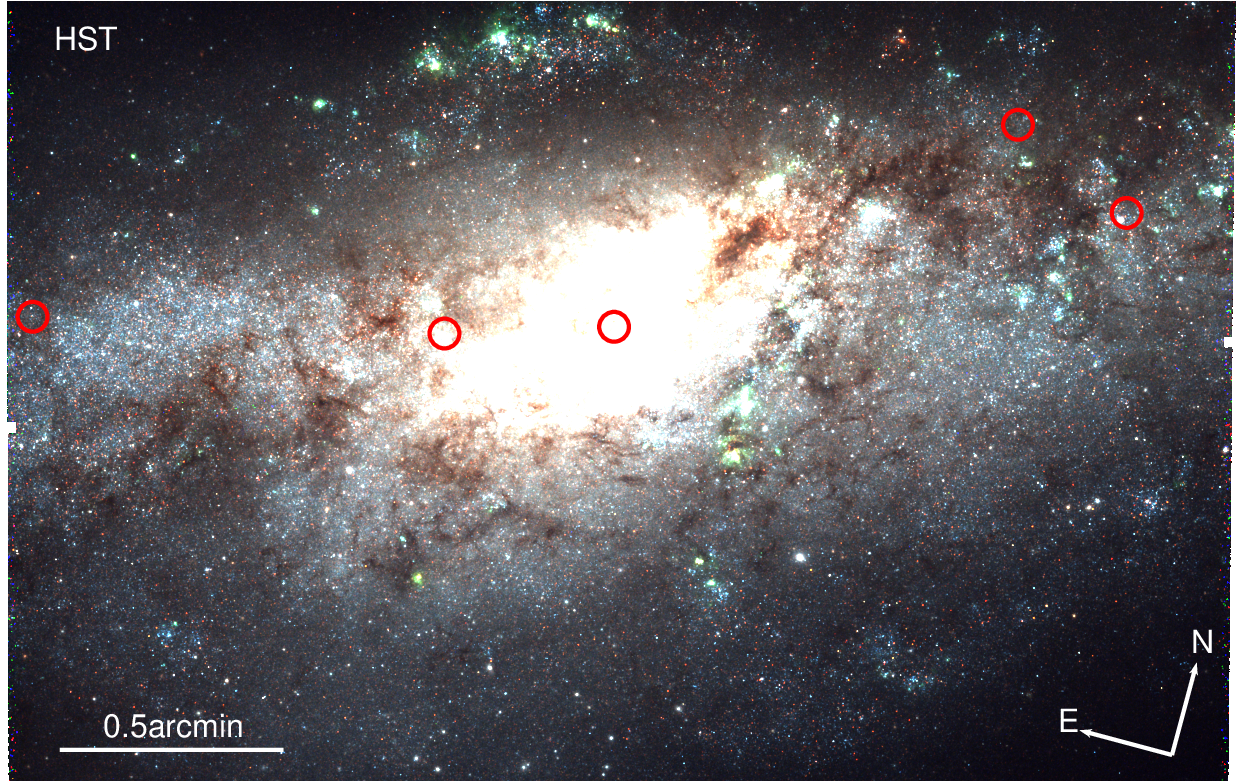}
\caption{%
NGC 4490 gökadasının Chandra (üst panel) ve HST (alt panel) görüntüleri. Chandra üç renk görüntüsü için kırmızı 0.3-1 keV, yeşil 1-2 keV ve mavi 2-10 keV enerji aralıkları kullanılmıştır. HST üç renk görüntüsü için kırmızı ACS/F814W, yeşil ACS/F555W ve mavi ACS/F438W görüntüleri kullanılmıştır.
}
\label{F:sinan}
\end{center}
\end{figure}

\begin{figure}
\begin{center}
\includegraphics[width=\columnwidth]{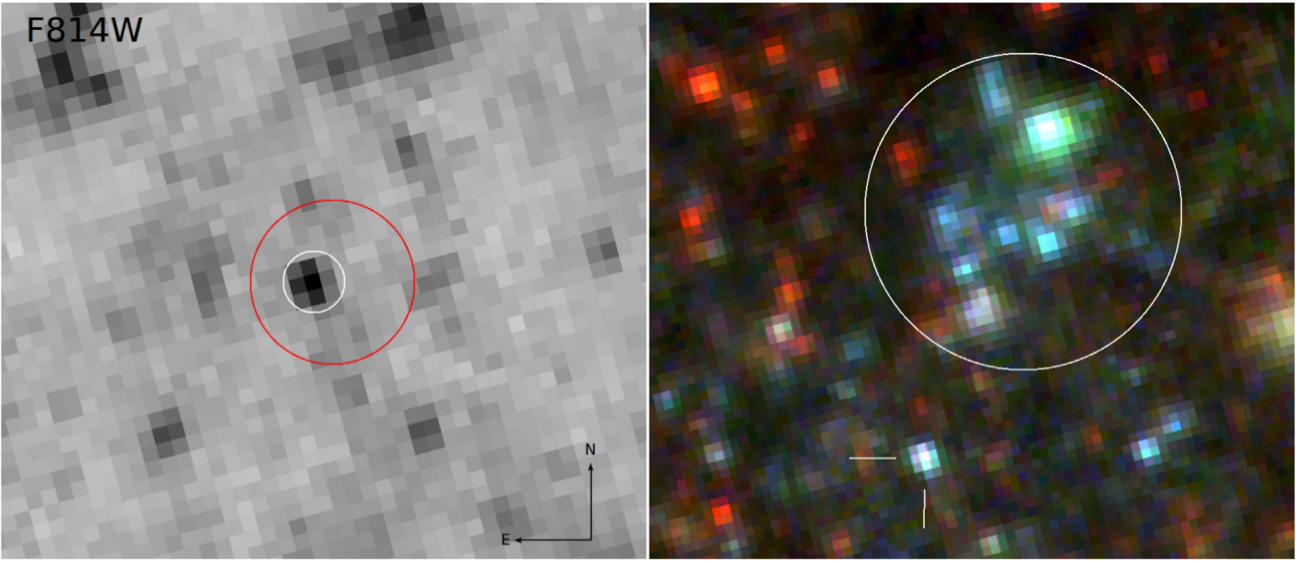}
\caption{NGC 4490 gökadasında APX X-4'ün HST / WFC3 F814W görüntüsü (sol) ve X-4 yakınlarındaki yıldız grubunun üç renk HST görüntüsü (sağ; kırmızı: F814W, yeşil: F555W, mavi: F438W). Sol paneldeki kırmızı daire yarıçapı 0$\arcsec$21 X-4'ün konumunu temsil eder. Yakınlaştırılan görüntünün alanı $\sim$1$\arcsec$.6$\times$1$\arcsec$.3'tür. Beyaz daire çapı 1$\arcsec$.4 olup, yıldız grubunun merkezinden X-4'e olan uzaklık $\sim$1$\arcsec$.2'dir.}
\label{F:4490}
\end{center}
\end{figure}

UZAYMER çalışma grubu ve Rusya Bilimler Akademisi Özel Astrofizik Gözlemevi (SAO RAS)'nden bir grup araştırmacı ile gerçekleştirilen "APX'lerin X-ışın ve optik özelliklerinin belirlenmesi" hakkında yaptığımız çalışmalar özetle:

NGC 4490/4485 etkileşen galaksi çiftinde yedi APX'ten beşinin olası optik karşılıkları belirlenmiştir \citep{2019ApJ...875...68A}. NGC 4490 galaksisinde bulunan APX'lerin konumlarının belirtildiği üç renk Chandra ve HST görüntüleri Şekil \ref{F:sinan}'de verilmiştir. HST arşiv verileri kullanılarak optik adayların mutlak parlaklıkları -3.5$-$-6.0 kadir arasında olduğu hesaplanmıştır. Optik adayların yaş ve kütle değerleri ve çevreleri ile olası bağlantılarını araştırmak için renk-kadir diyagramları kullanılmıştır. Adayların konumlarının, yakın yıldız grubu/kümesi içinde ya da yakınında olması durumunda olası oluşum ve fırlatma mekanizmaları araştırılmıştır (Şekil \ref{F:4490}). X-ışın arşiv verileri analiz edilerek, kaynak doğasının anlaşılmasında önemli olan uzun ve kısa süreli akı değişkenlikleri incelenmiştir. Ayrıca, incelenen çift sistemlerde periyodik bir sinyalin bulunamadığı durumlarda, sıkı cismin (olası bir kara delik) kütlesinin yıldız kütleli kara deliklere (10-15M$\odot$) işaret ettiği belirlenmiştir.
APX'lerin ışıma mekanizmasının anlaşılmasında önemli olan X-ışın ve optik özelliklerinin incelenmesi hakkında benzer çalışmalar; NGC 5474, NGC 3627 ve NGC 2500 galaksilerinde bulunan APX X-1 kaynakları \citep{2016ApJ...828..105A,2019MNRAS.488.5935A}, NGC 4258 galaksisinde bulunan APX X-3 ve X-6 kaynakları için yapılmıştır \citep{2016ApJ...828..105A, 2020MNRAS.499.2138A}. Bununla birlikte NGC 1316'da 2019 Chandra arşiv verileri kullanılarak APX X-7 olarak adlandırılan yeni bir kaynak tanımlamıştır. APX X-7 çoklu dalga boylarında incelenerek sistemdeki sıkı cisim kütlesinin $\sim$ 8M$\odot$ olabileceği belirtilmiştir. Ayrıca donör yıldızının tayfsal enerji dağılımı kara cisim modeli ile modellenerek olası bir geç tayf türü (M tipi süper dev) yıldız olduğu belirlenmiştir \citep{2020MNRAS.499.5682A}.

Bununla birlikte, UZAYMER çalışma grubunun APX'lerin incelenmesinin yanı sıra Samanyolu ötesi yıldız kümeleri ile X-ışın çiftleri arasındaki ilişkinin araştırılması konusundaki çalışmaları sürdürülmektedir.

\subsection{Ötegezegen Geçiş (Transit) Gözlemleri}
2020 yılında Nobel Fizik ödülünün verildiği ötegezegen araştırmaları \citep{1995Natur.378..355M}, gözlemsel astronominin popüler konularından biridir.
Ötegezegenler, güneş sistemi dışında bir yıldızın etrafında keşfedilen gezegen(ler) sistemidir. Yer ve uzay tabanlı teleskoplarla farklı yöntemler kullanılarak günümüze kadar $\sim$ 5000 ötegezegen sistemi keşfedilmiştir \citep{2015ARA&A..53..409W}. NASA'nın Kepler uydusu \citep{2010Sci...327..977B} ile gözlenen ötegezegenler geçiş yöntemi kullanılarak keşfedilmiştir. Güncel ötegezegen sayımlarına bakıldığında yine geçiş yöntemi ile bulunan gezegenler çoğunluktadır. Barınak yıldızına yakın bir yörüngede dolanan bu nesnelerin yörünge dönemleri birkaç gün mertebesindedir. 

Halen gözlemlerine devam eden ötegezegen keşif amaçlı çalışan teleskoplar için aday sistemlerdeki en önemli özelliklerden biri de zamanlamadır. Geçiş yaptığı bilinen ya da aday nesneler için bu geçişin başlangıç ve bitiş süresi ile geçiş orta zamanı hassas bir şekilde belirlenmelidir. Gözlem zamanının çok değerli olduğu uzay teleskoplarında gözlemi yapılacak bir sistem için tahsis edilecek gözlem süresinin hesaplanması gereklidir.
Bu nedenle yer tabanlı gözlemevlerinin bu tür gözlemler için kullanılması gereklidir.
UT50 ile gözlemi yapılan ötegezegen sistemlerinde amaç, barınak yıldızının önünden geçen gezegenin geçiş ortası zamanını hassas olarak belirlemenin yanı sıra varsa bu geçiş zaman değişimleri (TTV)'leri ve bu değişimlere neden olan etkileri ortaya çıkarmaktır.
Geçiş ortası zamanlarındaki TTV değişimlerine dair ölçümler UT50 gibi küçük çaplı teleskoplarla yapılabilmektedir \citep{2020PASP..132e4401Z}.

\begin{figure}
\begin{center}
\includegraphics[width=\columnwidth]{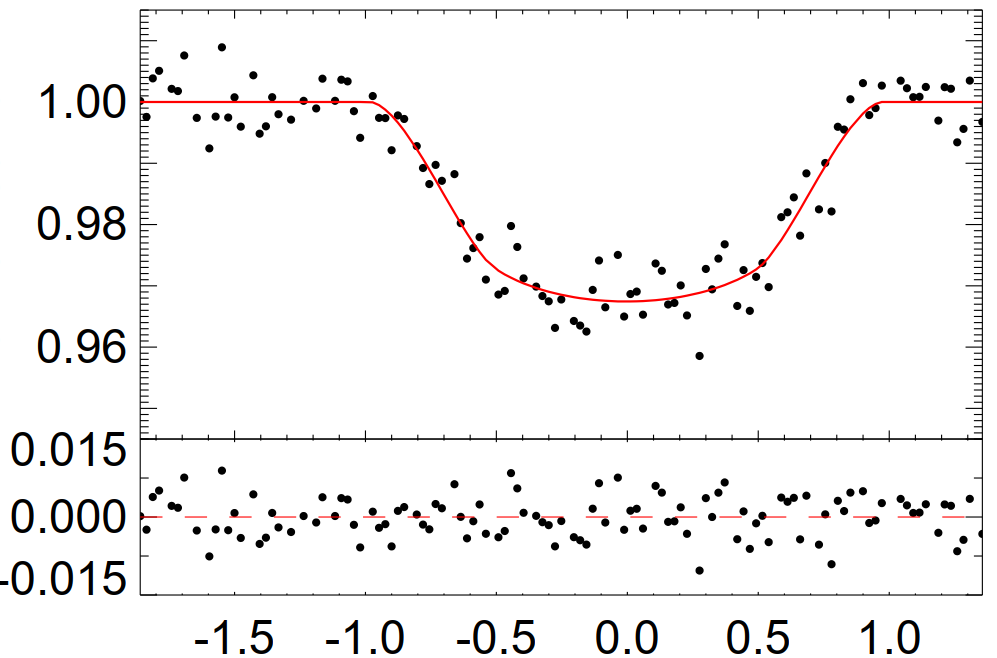}
\caption{%
UT50 ile yapılan WASP-52 b ötegezegen geçiş gözleminin EXOFAST programı ile yapılan model ışık eğrisi. Siyah noktalar gözlem verisi düz kırmızı çizgi ise EXOFAST programı ile üretilen modeli temsil eder (üst panel). Alt kısımda ise model ve alınan verilerin farkını gösteren kalıntılar (rezidü) yer almaktadır. Şekilde yatay eksen saat biriminde zamanı dikey eksen ise normalize edilmiş akıyı temsil etmektedir. 
}
\label{F:wasp52-20201023}
\end{center}
\end{figure}

UZAYMER'de gözlemi yapılan ötegezegen sistemlerinin ışık eğrisi çözümleri ve ön modelleme çalışmaları MaxIm DL, AstroImageJ \citep{2017AJ....153...77C} ve EXOFAST \citep{2013PASP..125...83E} programları ile yapılmaktadır. Gözlemi yapılan gezegen sistemlerden birisi de WASP-52'dir \citep{2013A&A...549A.134H}. Şekil \ref{F:wasp52-20201023}'da bu sistemin UT50 ile 23 Ekim 2020 gecesi alınan tipik bir geçiş gözlem verisinin EXOFAST ışık eğrisi model sonucu verilmiştir. Ayrıca bu gözlem verilerinden elde edilmiş olan bazı parametreler güncel çalışmalardan biri olan \cite{2021NewA...8301477S}'nin sonuçlarıyla karşılaştırmalı olarak Tablo \ref{T:tab_exo}'de sunulmuştur.

\subsection{Güneş Enerjisi Ölçümleri}

\begin{figure}
\begin{center}
\includegraphics[width=\columnwidth]{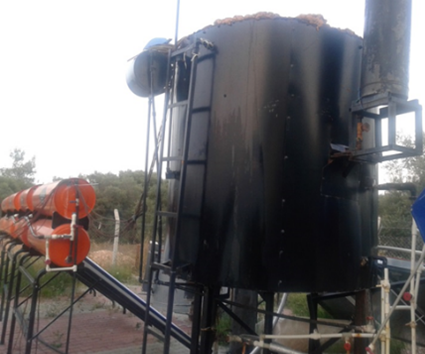}
\caption{%
Sağda, UZAYMER'deki Güneş havuzunu gösteren siyah silindirik yapı solda, güneş kolektörlerinin depoları olan turuncu yapılar görülmektedir.
}
\label{F:güneş-havuzu}
\end{center}
\end{figure}

\begin{figure}
\begin{center}
\includegraphics[width=\columnwidth]{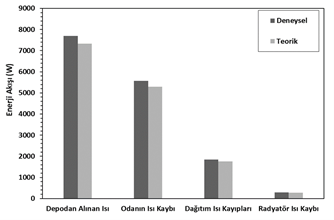}
\caption{%
UZAYMER'deki Güneş havuzu sisteminin teorik ve deneysel termal enerji dağılımı.
}
\label{F:termal-enerji}
\end{center}
\end{figure}

UZAYMER’de, güneş enerjisinden termal enerji üretmek için yalıtımlı silindirik model bir güneş havuzu ve vakum tüplü güneş kolektörleri kullanılmıştır (Şekil \ref{F:güneş-havuzu}). Deneysel ve teorik çalışmalar sistemin enerji dağılımlarının uyumlu olduğunu göstermektedir (Şekil \ref{F:termal-enerji}). Deneysel çalışmada sistemin deposundan çekilen ısı 7703 W, odaya aktarılan 5565 W dışarıdaki dağıtım kaybı 1846 W, radyatörün odaya veremediği ısı 292 W olmuştur. Teorik çalışmada ise sistemin deposundan çekilen ısı 7330 W, odaya aktarılan 5293 W dışarıdaki dağıtım kaybı 1759 W, radyatörün odaya veremediği 278 W ısı olmuştur. Bu sitem ile güneş enerjisinden elde edilen termal enerjiyle Merkezimizin ısı ve sıcak su ihtiyacının önemli bir bölümünü karşılayabileceği gösterilmiştir. Ayrıca, sistemin verimini etkileyen parametreler üzerinde çalışılmış ve enerji ve ekserji analizleri yapılmıştır \citep{Ayhan2014}. 

\begin{figure}
\begin{center}
\includegraphics[width=\columnwidth]{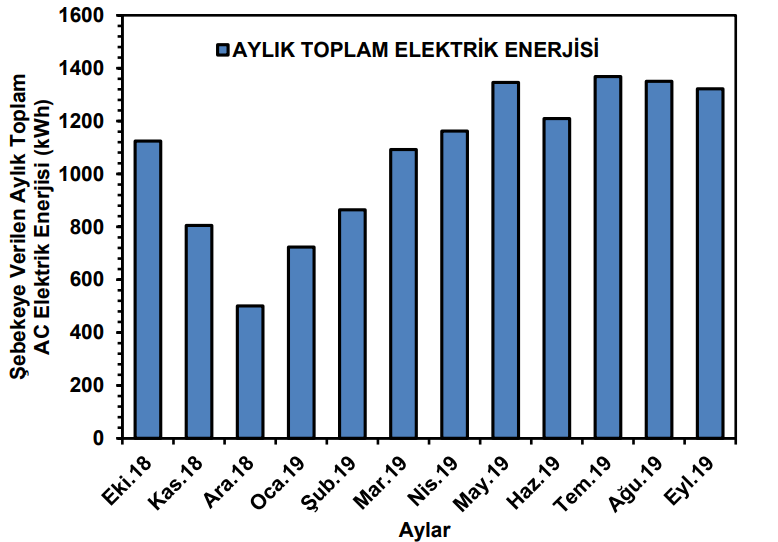}
\caption{%
UZAYMER'in mGES'den üretilen ve şebekeye verilen aylık toplam elektrik enerjisi (kWh).
}
\label{F:elektrik-enerjisi}
\end{center}
\end{figure}

UZAYMER’de, yenilenebilir enerji kaynaklarından biri olan güneş enerjisi ile merkezimizin elektrik ihtiyacını karşılamak ve gerekli ölçümleri yapmak için polikristal hücrelerden üretilen 32 adet güneş panelinden oluşan 8,6 kW kapasiteli şebekeye bağlı bir mini güneş enerjisi santrali (mGES) 2018 yılında kurulmuştur. Bu sistem Ç.Ü bünyesinde bu ölçekte kurulan ilk GES özelliğini taşımaktadır (Şekil \ref{F:map}'de ). Güneş enerjisi ile UZAYMER’in elektrik ihtiyacının yaklaşık \%80’i karşılanmaktadır. Ayrıca, verimi \%16 olan fotovoltaik panellerin performansını etkileyen faktörler araştırılmıştır. Fotovoltaik panellerin günlük elektrik üretimi, güneş radyasyonu, rüzgâr hızı, sıcaklık değerleri gibi parametreler ölçülmüştür. Panel sıcaklığı, hava sıcaklığı, nem, rüzgâr hızı, panel yüzey tozu, gölgeleme faktörü ve dönüştürücü (invertör) kayıpları panellerin verimini \%2-3 oranında etkilemektedir. Şekil \ref{F:elektrik-enerjisi}’de merkezimizdeki GES'in yıllık elektrik üretiminin aylara göre dağılımı verilmiştir. 
Şekil \ref{F:elektrik-enerjisi}’e göre, şebekeye verilen ortalama günlük elektrik miktarının, Aralık 2018'de 15 kWh en düşük, Temmuz 2019’da ise 44,136 kWh ile en yüksek değere ulaştığı görülmüştür \citep{ibrahim2020}. 

\section{UZAYMER'in Astrometeorolojik Durumu}
UZAYMER'e ait meterolojik parametreler, merkezin çatısında bulunan meteoroloji istasyonu verileri ile astronomik gözlemevleri için yer seçimi çalışması (AstroGIS;\citealp{2020MNRAS.493.1204A}) verileri kullanılarak incelenmiştir. Aşağıda kısaca meteoroloji istasyonu ve AstroGIS verileri hakkında bilgi verilmektedir.

\subsection{Meteoroloji İstasyonu}
\begin{figure*}
\begin{center}
    \begin{tabular}{@{}c@{}c@{}}
    \put(-220,0){\includegraphics[height=5.25cm]{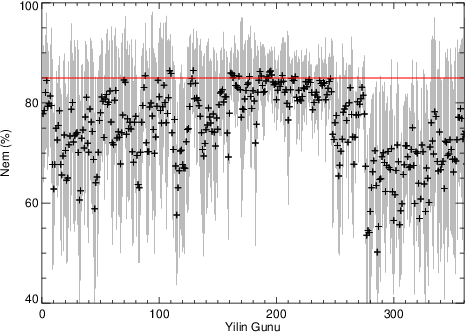}} \Large\color{red}\put(-25,138){a} &
    \put(0,0){\includegraphics[height=5.25cm]{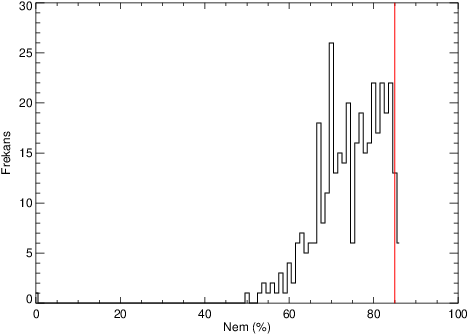}}
    \Large\color{red}\put(190,138){b} \\
    \put(-220,0){\includegraphics[height=5.25cm]{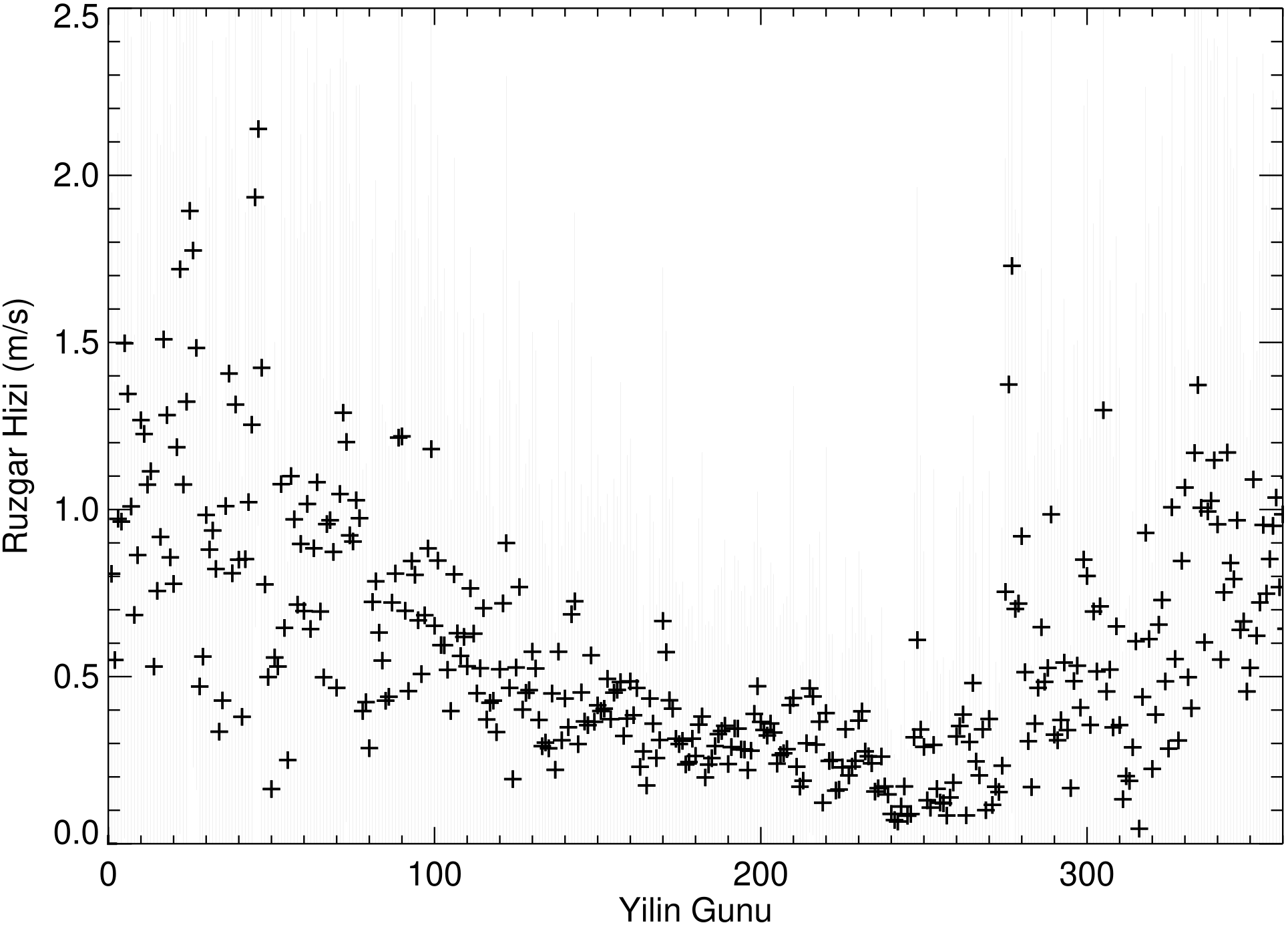}} 
    \Large\color{red}\put(-25,138){c} &
    \put(0,0){\includegraphics[height=5.25cm]{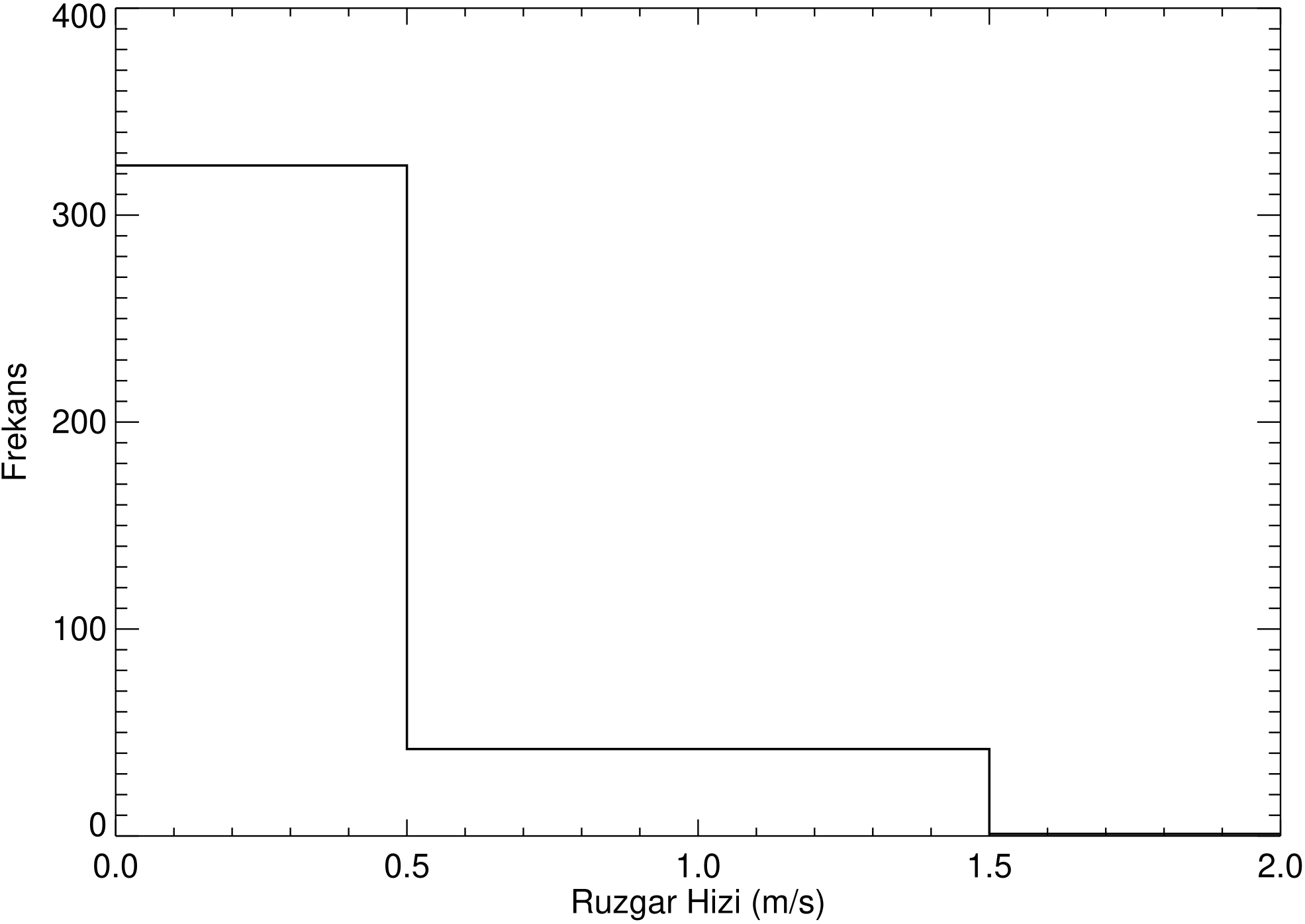}} 
     \Large\color{red}\put(190,138){d} \\
    \put(-220,0){\includegraphics[height=5.25cm]{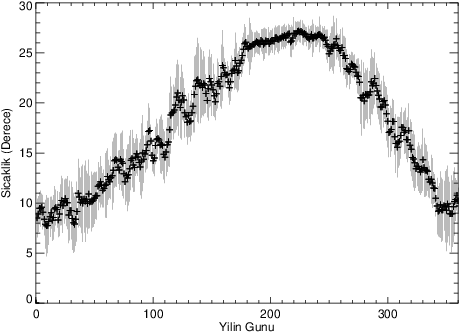}} 
    \Large\color{red}\put(-25,138){e} &
    \put(0,0){\includegraphics[height=5.25cm]{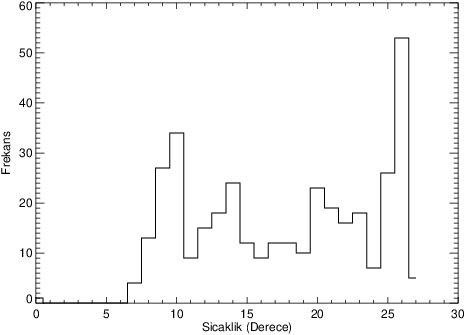}} 
     \Large\color{red}\put(190,138){f} \\
    \put(-220,0){\includegraphics[height=5.25cm]{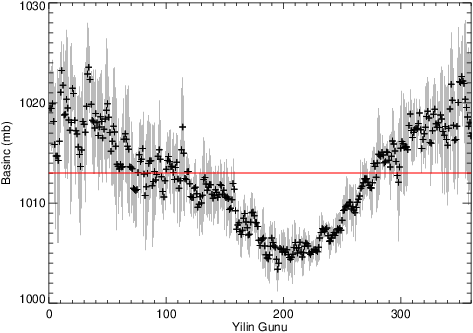}} 
    \Large\color{red}\put(-25,138){g} &
    \put(0,0){\includegraphics[height=5.25cm]{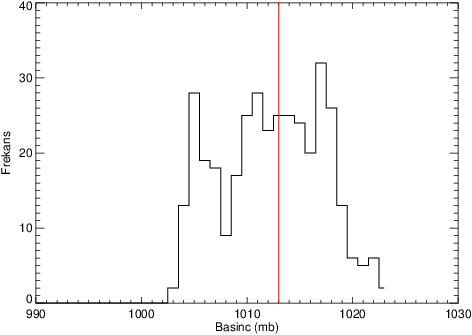}} 
     \Large\color{red}\put(190,138){h} \\
    \end{tabular}
\caption{%
UZAYMER için günlük ortalama Nem (a), rüzgâr hızı(c),  Sıcaklık(e) ve Basınç (g) grafikleri sol sütunda ve bu parametrelerin sıklık histogramları sağ sütunda (b, d, f ve h) verilmiştir. Grafiklerde gece hava durumunu incelemek amacıyla 20:00 - 03:00 arasında alınan veriler kullanılmıştır. Sol sütunda grafiklerdeki x-ekseni yılın gününü göstermektedir. + sembolü ilgili parametrenin ortalama değerini ve hata çubukları ise standart sapma değerini göstermektedir. Sağ sütunda grafiklerde y -ekseni gün sayısını göstermektedir. Grafiklerde kırmızı çizgiler Nem için \%85 ve Basınç için 1013 mb seviyesini göstermektedir. Rüzgâr limitlerinin üstünde gece sayısı yok iken nem değeri \%80 üstünde 106 gece ve \%85 üstünde 19 gece bulunmaktadır. Yaz aylarında daha az rüzgâr görülmekle birlikte nemin yüksek seyrettiği gözlenmektedir. Coğrafi konumdan beklendiği gibi yaz aylarında sıcaklıkta belirgin bir artış bulunmaktadır. Bununla birlikte basınç grafiğinde yaz aylarında alçak basınç görülmektedir.
}
\label{F:davis_time}
\end{center}
\end{figure*}

UZAYMER'de meteoroloji istasyonu olarak Davis Ventage Pro model ölçüm cihazı kullanılmaktadır. Meteoroloji istasyonu 19 Aralık 2012 tarihi itibariyle düzenli olarak nem, rüzgâr yönü ve rüzgâr şiddeti ile sıcaklık ve basınç ölçümleri yapmaktadır. Veri alma sıklığı 5 dakika olarak belirlenmiştir. Zaman zaman veri almadığı ya da teknik sorunların dışında toplam 568.925 ölçüm yapılmıştır. 20:00 - 03:00 saatleri aralığında 189.662 (toplam verinin \%33'ünü oluşturmaktadır) veri alınmış bu veriler gece verileri olarak işaretlenmiştir. Meteorolojik parametrelerin gece (gözlem yapılan saatler) verilerinden oluşturulan gecelik ortalama değerleri ve bunların histogram grafikleri Şekil \ref{F:davis_time}'de verilmiştir. Meteorolojik parametreler aşağıda listelenmiştir.

\subsubsection{Nem}
UZAYMER'de gözlemlerin durdurulmasına neden olan en önemli atmosferik parametre nemdir. Nem değeri \%85'i geçtiğinde gözlemler durdurulmaktadır. Toplam meteorolojik ölçüm sayısının \%24'üne karşılık gelen 138.708 ölçüm bu limitin üstündedir. Benzer şekilde gece ölçümlerinin \%32'sine karşılık gelen 60.512 ölçüm de bu limitin üstündedir. Ortalama nem değeri $\sim$ \%68 civarında hesaplanmıştır. Tüm gece boyunca yüksek nem (>\%85) nedeniyle gözlem yapılamayan gece sayısı yıllık ortalama 19'dur. Şekil \ref{F:davis_time}'de grafikleri incelendiğinde özellikle yaz aylarında gece boyunca nem değerlerinin ortalamanın (\%68) üzerinde seyrettiği görülmektedir.

\subsubsection{Rüzgâr Hızı ve Yönü}

Gözlemleri etkileyen bir diğer önemli atmosferik parametre ise rüzgâr hızıdır. UZAYMER'de rüzgâr hızı 11 m/s geçtiğinde gözlemler durdurulmaktadır. Meteorolojik ölçümlere göre, gözlem limitinin üzerinde sadece 60 ölçüm bulunmaktadır. Bu ölçümlerin çoğu 2 Kasım 2020 tarihinde alınmıştır. Ölçülen rüzgâr hızı verilerinden oluşturulan histogram Şekil \ref{F:davis_time}'de verilmiştir. Şekilde görüldüğü gibi gece rüzgâr değerlerinin oldukça düşük seyretmektedir. Yaz aylarında rüzgâr hızının azaldığı gözlenmiştir. Rüzgâr yönü ölçümlerinden hakim rüzgâr yönünün Kuzey-kuzeydoğu olduğu belirlenmiştir.

\subsubsection{Sıcaklık}

CCD'lerin yüksek sıcaklıklarda fazladan ek gürültü ürettikleri bilinmektedir \citep{1980PASP...92..233L}. UT50'de kullanılan CCD ortam sıcaklığının 60\degree C altına kadar soğuyabilmektedir. Dolayısıyla ortam sıcaklığı ne kadar düşük olursa gözlem kalitesi o kadar artmaktadır. Yıllık ortalama hava sıcaklığı gecelik 17.8 \degree C olarak hesaplanmıştır. Gecelik ortalama sıcaklık grafiği ve histogramı Şekil \ref{F:davis_time}'de verilmiştir. Gecelik sıcaklık değerlerinin 6 ile 26 \degree C arasında dağıldığı ve yaz aylarında bu değerlerin arttığı gözlenmiştir.

\subsubsection{Hava Basıncı}

1013 mb basınç normal basınç (1 Atm) olarak kabul edilmektedir. Bu değerin altı düşük basınç veya açık gökyüzü olarak tanımlanmaktadır. Ortalama değer geceleri 1012.8 mb olarak ölçülmüştür. Gecelik ortalama basıncın grafiği ve histogramı Şekil \ref{F:davis_time}'de verilmiştir. Özellikle yaz aylarında düşük basınç değerleri gözlenmiştir.

\subsection{AstroGIS}
Tüm dünyada gözlemevleri için yer seçim çalışmasında güncel ve yüksek çözünürlüklü AstroGIS \footnote{astrogis.org.tr} veri tabanındaki veriler kullanılmaktadır. AstroGIS veri tabanı çoğunlukla uydu ve model verilerinden üretildiğinden küresel yer seçimi çalışmaları için uygun olmasına rağmen lokal yer seçimler için de kullanılabilmektedir. UZAYMER'in konumu kullanılarak AstroGIS verilerinden elde edilen atmosferik parametrelere ait aylık ve yıllık değişimler Şekil \ref{F:fig}'te ve yıllık istatistikler Tablo \ref{T:tab}'de verilmiştir. UZAYMER'in AstroGIS verilerinden üretilen Astronomik Uygunluk İndeksi (SIAS)'ın A kategorisinde bulunan 2141 gözlemevi içerisinde 972. sırada bulunduğu belirlenmiştir \citep{2020MNRAS.493.1204A}.

\subsubsection{Bulutluluk}

Astronomik gözlemleri en çok etkileyen faktörlerden biri de bulutluluktur. Bu nedenle UZAYMER'in son 20 yıldaki bulutluluk grafiği aylık ve yıllık olarak çıkarılmıştır ve Şekil \ref{F:fig}'de verilmiştir. Bulutluluğun, Ocak ve Nisan aylarında maksimum iken Eylül ve Kasım aylarında minimum olduğu gözlenmiştir. Yıllık ortalama bulutluluk grafiğinde 2017 yılında belirgin bir düşüş ve onu takip eden 2018 yılında bir artış görülmektedir. Tablo \ref{T:tab}'de yıllık ortalamalardan elde edilen istatistiksel sonuçlar verilmiştir. Yıllık ortalama minimum değer \%53.17, maksimum değer \%71.12 ve ortalama bulutluluk değeri \%61.97 olarak belirlenmiştir. Bu durumda UZAYMER'de açık gece sayısı yılda ortalama 226 gece olmaktadır.

\subsubsection{Işık Kirliliği}

\begin{figure}
\begin{center}
    \includegraphics[height=5.25cm]{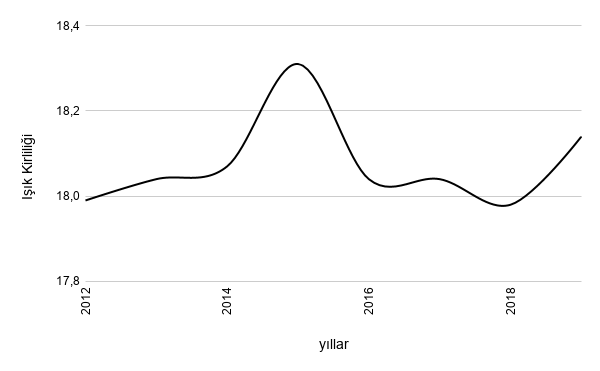}
    \includegraphics[height=5.25cm]{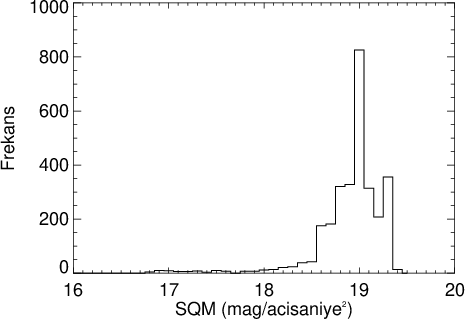}
\caption{%
UZAYMER’in uydu verilerinden elde edilen yıllık ortalama ışık kirliliği değerlerinin değişimi (üstte) ve SQM ölçümlerinden oluşturulan histogram (altta) gösterilmiştir. Histogram grafiğinde 19 mag/arcsec$^{^2}$ civarında bir kümelenme görülmektedir. Bortle ölçeğine göre UZAYMER şehir merkezinden uzak yerleşimdeki gökyüzü ve dolunay sınıfına girmektedir.
}
\label{F:lp}
\end{center}
\end{figure}

Astronomik ölçümleri etkileyen bir diğer önemli faktör ışık kirliliğidir. UZAYMER'in ışık kirliliği uydu verileri kullanılarak ölçülmüştür \citep{2020MNRAS.493.1204A}.
Uydu verilerinden elde edilen sonuçlara göre ortalama ışık ölçümü değeri 18.1 kadir/açı saniye$^{^2}$ (mag/arcsec$^{^2}$), minimum ve maksimum değerler ise 18.0 ve 18.3 mag/arcsec$^{^2}$ olarak belirlenmiştir. 
Tablo \ref{F:lp}'de ölçülen yıllık ışık ölçüm değerlerinin grafiği verilmiştir.
Uydu verilerine ek olarak Gökyüzü Kalite Ölçeri (Sky Quality Meter - SQM) SQM-LE/LU cihazıyla ölçümler yapılmıştır. SQM gökyüzü parlaklığını mag/arcsec$^{^2}$ biriminde ölçmektedir. 21 Temmuz 2020 ile 15 Ocak 2021 tarihleri arasında UZAYMER'de baş ucu doğrultusunda yapılan 2952 ölçümde ortalama, minimum ve maksimum değerler sırasıyla 18.9, 16.8 ve 19.4 mag/arcsec$^{^2}$ olarak ölçülmüştür.

Bu ölçümler, Güneş'in ufkun (-18 \degree'nin) altında (astronomik tan) ve Ay'ın ufkun (0 \degree'nin) altında olduğu zamanlarda alınmıştır. SQM ölçümlerinden oluşturulan histogram Şekil \ref{F:lp}'de verilmiştir. Bu ölçümler Bortle ölçeğine göre \citep{Bortle2001} şu şekilde değerlendirilebilir: 21.99 > mükemmel gökyüzü, 21.89-21.99: karanlık gökyüzü, 21.69-21.89: kırsal alanlar, 21.25-21.69: kırsal alan sınırları, 20.49-21.25: kenar mahalle sınırları, 19.50-20.49: kenar mahalle gökyüzü, 18.38-19.50: şehir merkezinden uzak yerleşimdeki gökyüzü ve dolunay, < 18.38: şehir merkezi. Bu ölçeklendirmeye göre UZAYMER'in konumu şehir merkezinden uzak yerleşimdeki gökyüzü ve dolunay olarak sınıflandırılabilir.

\subsubsection{Aerosol}

Aerosol gaz halde bulunan herhangi bir kütle içerisinde asılı olan katı ve sıvı parçacıkların meydana getirdiği ince karışm olarak tanımlanır. Gökyüzündeki aerosoller havanın ağırlaşmasına ve opaklığın kaybolmasına neden olur. Havadaki aerosol miktarı arttıkça atmosferik görüş azalmaktadır. Aerosollerin bulutlar ve diğer atmosferik tabakalarla etkileşimi meteorolojik olaylarda katalizör olmasına rağmen, astronomik ölçümlerde istenmeyen bir durumdur \citep{varela2008}. Şekil \ref{F:fig}'de UZAYMER'e ait aylık ve yıllık ortalama aerosol optik derinliğin grafiği verilmiştir. Aylık ortalamalarda Ağustos - Eylül aylarında maksimum seviyeye ulaşan aerosol miktarı Nisan - Temmuz arası sabitken Aralık-Ocak aylarında minimum seviyededir. Yıllık ortalamalar grafiğinde ise 2008, 2014 ve 2018 yıllarında artış gözlenmektedir. Tablo \ref{T:tab}'de yıllık ortalamalardan elde edilen istatistiksel sonuçlara göre ortalama, minimum ve maksimum aerosol optik derinlik değerleri sırasıyla 2.34, 1.94 ve 3.34'dür. 
 
\subsubsection{Su Buharı}

\begin{figure*}
    \centering
    \begin{tabular}{@{}c@{}c@{}}
    \put(-220,0){\includegraphics[height=4.25cm]{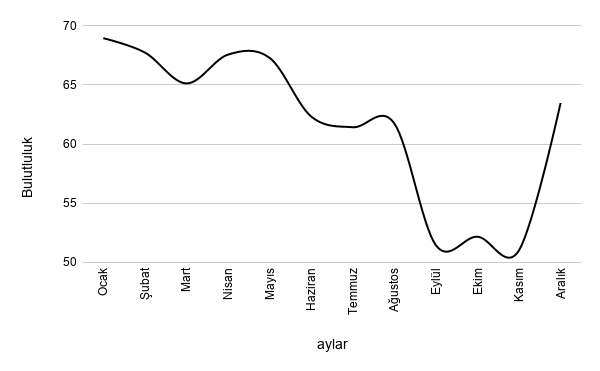}} 
    \Large\color{red}\put(-40,105){a} &
    \put(0,0){\includegraphics[height=4.25cm]{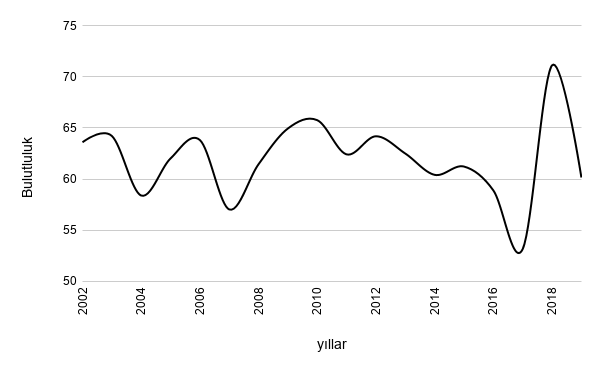}}      
    \Large\color{red}\put(180,105){b} \\
    \put(-220,0){\includegraphics[height=4.25cm]{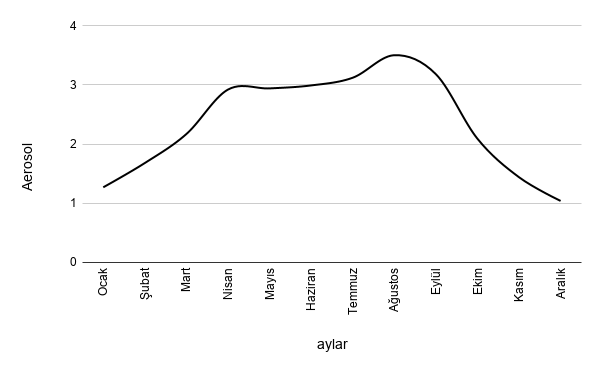}}  
    \Large\color{red}\put(-40,105){c} &
    \put(0,0){\includegraphics[height=4.25cm]{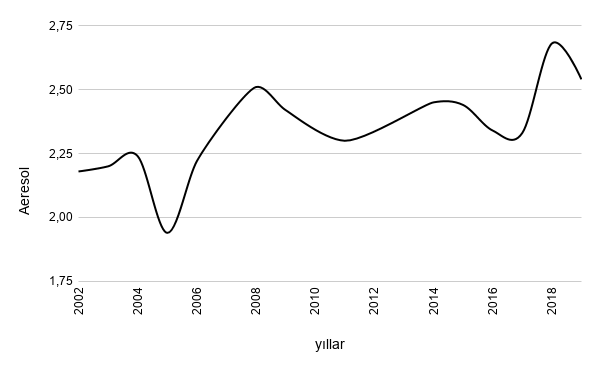}}  
    \Large\color{red}\put(180,105){d} \\
    \put(-220,0){\includegraphics[height=4.25cm]{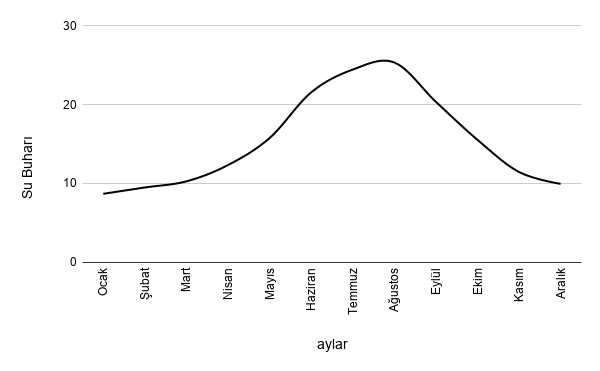}}     
    \Large\color{red}\put(-40,105){e} &
    \put(0,0){\includegraphics[height=4.25cm]{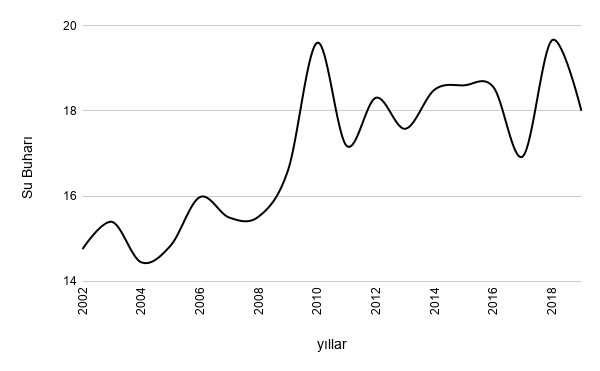}}     
    \Large\color{red}\put(180,105){f} \\
    \put(-220,0){\includegraphics[height=4.25cm]{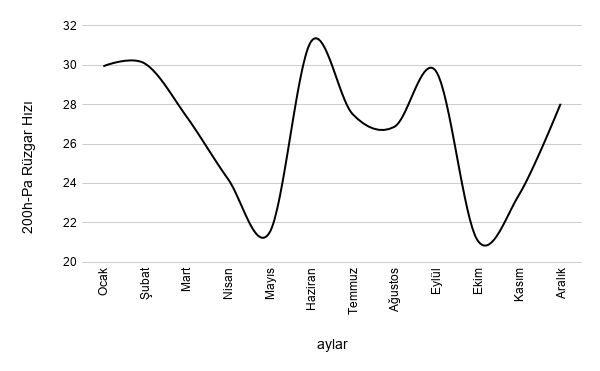}} 
    \Large\color{red}\put(-40,105){g} &
    \put(0,0){\includegraphics[height=4.25cm]{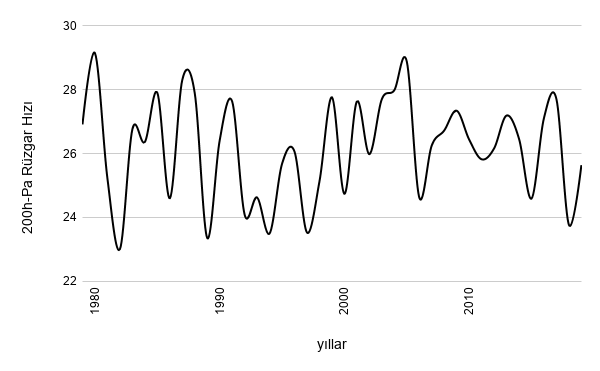}} 
    \Large\color{red}\put(180,105){h} \\
    \put(-220,0){\includegraphics[height=4.25cm]{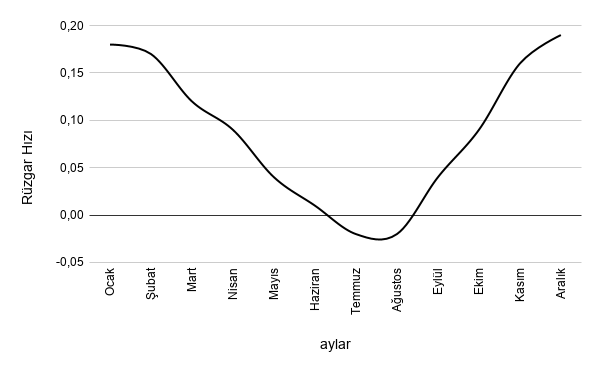}}    
    \Large\color{red}\put(-40,105){i} &
    \put(0,0){\includegraphics[height=4.25cm]{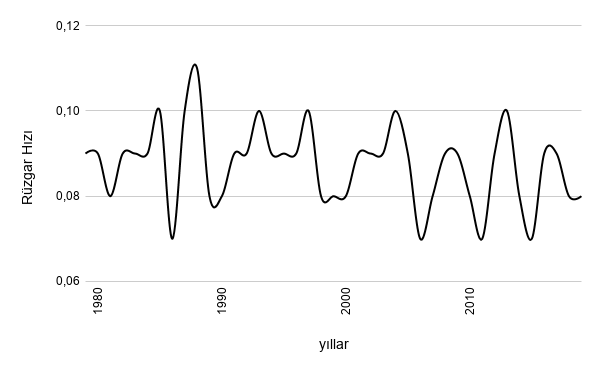}}     
    \Large\color{red}\put(180,105){j} \\
    \end{tabular}
\caption{%
UZAYMER’in coğrafi konumu ile ilişkili bulutluluk (a), aerosol optik derinlik (c), yoğuşabilir su buharı (e), 200h-Pa basınç seviyesindeki rüzgâr hızı (g) ve rüzgâr hızı (j) katmanlarının aylık (sol tarafta) ve yıllık (sağ tarafta-b, d, f, h ve j) değişimleri.%
}
\label{F:fig}
\end{figure*}

Su buharı, özellikle kırmızı ötesi dalga boyunda gözlemlerin yapıldığı astronomik gözlemevleri için önemli bir parametredir \citep{perez2015}. Su buharı ölçümleri 1 mm çapında silindirik sütun içerisindeki yükseklik (mm) olarak belirlenir ve yer yüzeyine yakın bölgedeki nem değeri ile ilişkilidir. Yüksek nem, bu tür ortamlarda bulunan gözlem araçları üzerinde oksitlenme vb. nedenlerle negatif bir etkiye sahiptir. UZAYMER'e ait aylık ve yıllık ortalama yoğuşabilir su buharı grafiği Şekil \ref{F:fig}'te verilmiştir. Yaz aylarında su buharı en yüksek değere ulaşır, kış aylarında ise bu değer daha düşüktür. Yıllık ortalamalar grafiği incelendiğinde ise 2002 yılından 2019 yılına kadar olan süre zarfında 2008 sonrasında gerçekleşen artış dikkat çekmektedir. Tablo \ref{T:tab}'de verilen su buharına ait R-kare değeri diğer katmanlara göre en yüksek R-kare değeridir. R-kare değeri 0 ile 1 aralığında olup 1'e yakın değerler veriler ile modelin uyumunu göstermektedir. Şekil \ref{F:fig}'te yıllık ortalamalar grafiğindeki su buharındaki artışı göstermektedir. istatistiksel sonuçlara göre ortalama, minimum ve maksimum yoğuşabilir su buharı değeri sırasıyla 17.0 mm, 14.5 mm ve 19.66 mm'dir. Türkiye'de yoğuşabilir su buharı değeri 0.4-2.0 aralığında değişmektedir \citep{2015ExA....39..547A}. Buna göre UZAYMER'in oldukça yüksek yoğuşabilir su buharına sahip olduğu görülmektedir.

\subsubsection{200hpa Rüzgâr Hızı}

200 hectopascal (h-Pa) rüzgâr hızı katmanı türbülans yaratması nedeniyle astronomik görüşü düşürmektedir \citep{2019MNRAS.482.4941H}. Bu nedenle UZAYMER'e ait aylık ve yıllık ortalama 200h-Pa basınç seviyesindeki rüzgâr hızı araştırılmış ve değerler Şekil \ref{F:fig}'de verilmiştir. Mayıs ve Ekim aylarında en düşük olan rüzgâr hızı Şubat ve Mayıs arasında en yüksek seviyededir. Yıllık ortalamalarda ise genel bir artış ve azalış eğilimi görülmemektedir.

\subsubsection{Dikey Rüzgâr Hızı}

Gözlemevi seviyesinde rüzgârın coğrafi şekillere çarptıktan sonra türbülans yaratması nedeniyle astonomik görüş etkilenmektedir. Ayrıca rüzgâr hızının 11 m/s'den yüksek olduğu durumlarda teleskop ve kubbe üzerinde oluşan yük nedeniyle titreşim yaratabilmektedir \citep{Liu_2020}. Bu nedenle UZAYMER'e ait aylık ve yıllık ortalama rüzgâr hızı grafiği Şekil \ref{F:fig}'de verilmiştir. Yaz aylarında düşük olan rüzgâr hızı aralık ve ocak aylarında yüksektir. Yıllık ortalamalarda bir artış veya azalış eğilimi görülmemektedir.

\section{Sonuç ve Öneriler}
Ülkemizde 50 cm ve üzeri aynı çaplı teleskop içeren sayılı gözlemevlerinden biri olan UZAYMER'de 1991 yılından bu yana astronomi-astrofizik, güneş enerjisi ve uzaktan algılama alanlarında bilimsel araştırmalar sürdürülmektedir. Bu çalışmada UZAYMER'in faaliyet gösterdiği araştırmaların yanı sıra astrometeorolojik koşulları ve altyapısı ile ilgili ölçümler sunulmuştur.
UZAYMER'de bulunan teleskoplar ve meteoroloji istasyonu profesyonel ve eğitim amaçlı gözlemler ve ölçümler yapmaya elverişlidir. 
\begin{itemize}
\item Gözlemevinin astronomik görüşü UT50'deki CCD görüntüleri üzerinden 3.5 arcsec olarak belirlenmiştir. 
\item 120 s poz süresindeki fotometrik sonuçlara göre 100 S/G için limit parlaklıklar B, V, R ve I filtrelerinde sırasıyla 12.9, 12.4, 12.0 ve 11.7 kadir olarak belirlenmiştir.
\item UZAYMER altyapısında yapılan iyileştirmeler sayesinde UT50 uzaktan kontrol edilebilmektedir. 
\item UT50'nin ilk ışığından bu yana yaklaşık 130 farklı astronomik kaynak gözlenmiş ve önemli bir veri arşivi ($\sim$65000 FITS formatında veri) oluşmuştur. 
\item UZAYMER çalışma grubu, TÜBİTAK projelerinin desteğiyle APX’lerin X-ışın ve optik özelliklerinin incelenmesinin yanı sıra galaksi ötesi yıldız kümeleri ile X-ışın çiftleri arasındaki ilişkinin araştırılması konusunda çalışmalarını sürdürmektedir.
\item UT50'nin gözlem zamanının \% 60'ı ötezgezegenlere ayrılmıştır. Bu çalışmada WASP-52 b ötegezegen gözlemlerinin sonuçları verilmiştir. 
\item UZAYMER'in çatısına 2019 yılında mGES kurulmuştur. Merkezin gündüz elektrik tüketim ihtiyacının maksimum \%80'i mGES'ten sağlanmaktadır.
\item UZAYMER'in astrometeorolojik durumu özellikle gece alınan uydu ve meteoroloji istasyonu verileri ile incelenmiştir. Buna göre;
Gözlemleri en çok etkileyen faktör nem değerinin yüksekliğidir (Ortalama=\%68). Yıllık ortalama 19 gece \%85'den büyük olmaktadır.
Hakim rüzgâr yönü Kuzey-kuzeydoğu'dur. Limitleri aşan (>11 m/s) rüzgâr neredeyse hiç yoktur.
Yıllık ortalama açık gece sayısı 225'tir. En az bulutluluk Eylül-Kasım ayları arasındadır.
Su buharı seviyesi son 20 yılda artış eğilimindedir.
\item UZAYMER'de ışık kirliliği seviyesi uydu verileri ve SQM ölçümleri ile ortalama 19 mag/arcsec$^{^2}$ olarak belirlenmiştir. 
\end{itemize}

UZAYMER, \cite{2020MNRAS.493.1204A} tarafından geliştirilen Astronomik Gözlemevleri için Uygunluk İndisi (SIAS) ölçeğinin A kategorisinde 2141 gözlemevi içerisinde 0.65 değeri ile 972. sırada bulunmaktadır. Bu değer UZAYMER'in Dünya ortalamasının üzerinde olduğunu göstermektedir. Tüm ölçümler ve hesaplamalar dikkate alındığında UZAYMER'in küçük-orta ölçekli ulusal ve uluslararası ortaklı bilimsel çalışmalar için yeterli altyapıya sahip olduğu değerlendirilmektedir.

\section{Teşekkür}
Bu çalışma, AstroGIS verilerinin üretildiği 117F309 nolu TÜBİTAK ve FYL-2019-11834 numaralı Çukurova Üniversitesi Bilimsel Araştırmalar Birimi (BAP) projeleri tarafından desteklenmiştir. APX'ler ile ilgili çalışmalar 117F115 numaralı TÜBİTAK projesi tarafından desteklenmiştir. 118F042 numaralı TÜBİTAK projesi tarafından desteklenen ötegezegen çalışmaları için Özgür Baştürk'e teşekkür ederiz. SQM-LE/LU 4580 seri numaralı alet ile yapılan ölçümler Çağ Üniversitesi BAP birimi tarafından desteklenen proje kapsamında alınmıştır. Şekil 1'e katkısı için Murat Beyazit'a teşekkür ederiz. AstroGIS verileri için gruba (https://www.astrogis.org/about) teşekkür ederiz.

\begin{table*}
    \caption{UZAYMER'den alınan verilerden oluşturulan V, R ve I filtrelerinde birinci dereceden gecelik sönümleme katsayıları.}
    \centering
    \begin{tabular}{cc}
    \hline \hline
    Sönümleme Katsayısı & Değer \\
    \hline 
$k_v$ & {0.2937$\pm$0.021} \\
$k_r$ & {0.1904$\pm$0.020}	\\
$k_i$ & {0.1225$\pm$0.019}	\\	
    \hline\hline 
    \end{tabular}
    \label{T:tab_katsayılar}
\end{table*}

\begin{table*}
    \caption{UT50 ile 2 Ekim - 4 Aralık 2020 tarihleri arasında 6 farklı gecede alınan WASP-52 gözlem verilerinin EXOFAST programından elde edilen model parametreleri ve \citep{2021NewA...8301477S} ile karşılaştırılması.}
    \centering
    \begin{tabular}{lcccc}
    \hline \hline
    Parametre & Simge & Birim & UT50 & SA21 \\
    \hline
    Yıldız Parametreleri \\
Kütle	& $M_{\ast}$ &	$M_{\odot}$ &0.816$\pm$0.0006	&	0.844$\pm$0.038   \\
Yarıçap & $R_{\ast}$	&$R_{\odot}$	&0.767$\pm$0.0003	&	0.836$\pm$0.013	\\
Işıtma & $L_{\ast}$	&$L_{\odot}$	&0.330$\pm$0.001	&	0.399$\pm$0.013	\\
Yoğunluk & ${\uprho}_\ast $ & g/cm$^{3}$	&2.551$\pm$0.0009	&	2.040$\pm$0.175	\\
Yüzey çekim ivmesi & $\log{g}$	& cm/s$^{2}$	&4.580$\pm$0.00001	&	4.520$\pm$0.022	\\
Etkin Sıcaklık & $T_{eff}$		&Kelvin	&4999$\pm$3.20	&	5017$\pm$41	\\
Metal Bolluğu & $[Fe/H]$	&	&0.029$\pm$0.002	&	0.130$\pm$0.110	\\
\hline
    Gezegen Parametreleri \\
Dışmerkezlik	& $e$ &	&0.053$\pm$0.048	&	0.050$\pm$0.026	\\
Yörünge Dönemi & $P$	& Gün	&1.749$\pm$2.43$\times$10$^{-7}$	&	1.749$\pm$1.37$\times$10$^{-7}$	\\
Yarı-büyük Eksen & $a$	& AB	&0.027$\pm$6.52$\times$10$^{-6}$	&	0.026$\pm$0.0004	\\
Yarıçap & $R_{P}$ 	&$R_J$	&1.279$\pm$0.081	&	1.322$\pm$0.0026	\\
Denge Sıcaklığı & $ T_{eq} $	&	Kelvin&1296$\pm$0.906	&	1349$\pm$13.5	\\
\hline
    Geçiş Parametreleri \\
Geçiş orta-zamanı	& BJD-TDB &	&2459146.260508$\pm$0.000475	&	2455793.680953$\pm$0.000149	\\
Yarıçap Oranı & $R_{P}/R_{\ast}$	&	&0.171$\pm$0.011	&	0.162$\pm$0.002	\\
Yarı-büyük Eksen Oranı & $a/R_{\ast}$	&	&7.444$\pm$0.001	&	6.910$\pm$0.190	\\
Yörünge Eğimi & $i$	& Derece	&85.33$\pm$0.535	&	84.9$\pm$0.240	\\
Etki Parametresi & $b$	&	&0.639$\pm$0.070	&	0.590$\pm$0.020	\\
Geçiş Derinliği & $\updelta$	&ppt	&0.029$\pm$0.0004	&	0.026$\pm$0.0006	\\
Geçiş Süresi & $ T_{14} $	&Gün	&0.079$\pm$0.0003	&	0.0779$\pm$0.0006	\\
    \hline\hline 
    \end{tabular}
    \label{T:tab_exo}
\end{table*}

\begin{table*}
    \caption{UZAYMER’in coğrafi konumunun bulutluluk, yoğuşabilir su buharı, aerosol optik derinlik, rüzgâr hızı, 200h-Pa basınç seviyesindeki rüzgâr hızı katmanlarının ölçülen minimum, ortalama, sigma, maksimum, R-kare ve eğim değerleri.}
    \centering
    \begin{tabular}{lccccccc}
    \hline \hline
    Katmanlar & Birim & Minimum & Ortalama & Std. Sapma & Maksimum & R$^{2}$  & Eğim \\
    \hline 
Bulutluluk	& {\%} &	53.17	&	61.97	&	3.76	&	71.12	&	0.01	&	-0.06	\\
Işık Kirliliği & {mag/arcsec$^{2}$}	&	17.98	&	18.07	&	0.09	&	18.31	&	0.00	&	0.01	\\
Yoğuşabilir Su Buharı & {mm}	&	14.45	&	17.00	&	1.64	&	19.66	&	0.66	&	0.26	\\
Aerosol Optik Derinlik & {değer}	&	1.94	&	2.34	&	0.18	&	3.34	&	0.51	&	0.02	\\
200hPA da Dikey Rüzgâr & {m s$^{-1}$}	&	23.01	&	26.16	&	1.58	&	29.14	&	0.00	&	0.00	\\
Rüzgâr Hızı & {Pa s$^{-1}$}	&	0.07	&	0.09	&	0.01	&	0.11	&	0.07	&	0.00	\\
    \hline\hline 
    \end{tabular}
    \label{T:tab}
\end{table*}

\bibliographystyle{tjaa}
\bibliography{UT50}

\bsp	
\label{lastpage}
\end{document}